\documentclass[11pt,a4paper]{article}

\usepackage[utf8]{inputenc}
\usepackage[T1]{fontenc}
\usepackage{lmodern}

\usepackage{amsmath,amssymb,amsthm}
\usepackage{algorithm}
\usepackage{algorithmic}

\usepackage{geometry}
\usepackage{microtype}

\usepackage{graphicx}
\usepackage{subcaption}
\usepackage{placeins}
\usepackage{authblk}
\usepackage{booktabs}
\usepackage{tabularx}
\usepackage{array}
\usepackage{float}
\usepackage[dvipsnames]{xcolor}

\usepackage{tikz}
\usetikzlibrary{positioning,arrows.meta,calc,fit,backgrounds}

\usepackage[round,authoryear]{natbib}

\usepackage{hyperref}

\long\def\added#1{#1}

\newcommand{\todo}[1]{}

\title{A Geometry-Aware Residual Correction of Hagan's SABR Implied Volatility Formula}

\author[1,2]{Adil Reghai}
\author[2]{Lama Tarsissi}
\author[3,4]{G\'erard Biau}
\author[1,5]{Alex Lipton}

\affil[1]{Abu Dhabi Investment Authority (ADIA)}
\affil[2]{SAFIR, Sorbonne University Abu Dhabi (SUAD)}
\affil[3]{Sorbonne Universit\'e, LPSM, Paris}
\affil[4]{Institut Universitaire de France}
\affil[5]{ADIA Lab}

\date{\today}

\begin{document}
\maketitle

\begin{abstract}
This paper proposes a hybrid methodology to improve the approximation of SABR
(Stochastic Alpha Beta Rho) implied volatility. The residual-learning idea, on
which our framework builds, is not new: it goes back to control-variate and
asymptotic-correction approaches in the neural-network SABR literature. Our contribution
introduces two departures from this line. First, we residualise Hagan's formula
multiplicatively rather than additively, which keeps the correction on the same
scale as the analytical baseline and preserves positivity. Second, we augment the network input with
geometry-aware features derived from the intrinsic Riemannian metric of the SABR
dynamics, in addition to the raw SABR parameters.

Unlike approaches that fully replace analytical formulas with black-box models, the
proposed framework preserves the analytical backbone of the model. Numerical
experiments, including two controlled ablations that isolate the effect of the
multiplicative form and of the geometric features, together with a Monte Carlo
convergence study and a static-arbitrage diagnostic on the learned surface, show that
both ingredients deliver measurable and structurally consistent improvements over the
residual-only baseline. Because the correction remains lightweight and structurally
consistent with the underlying model, the framework is well suited for real-time
pricing and calibration in practical trading environments.
\end{abstract}

\newpage

\section{Introduction}

The accurate approximation of implied volatilities under the SABR (Stochastic Alpha Beta Rho)
model has long been a central problem in quantitative finance, owing to the analytical
elegance of the model and its pervasive use in global derivatives markets. The closed-form
expansion introduced by \citet{Hagan2002} rapidly became the industry standard due to its
computational speed, smoothness, and ease of calibration.

Despite its widespread adoption, the Hagan formula is only an approximation. Its limitations are
well documented: systematic biases appear away from the at-the-money (ATM) region, both in the
in-the-money (ITM) region ($K < F_0$) and in the out-of-the-money (OTM) region ($K > F_0$).
Throughout the paper, we distinguish between the strict ATM point ($K = F_0$) and the
ATM region, which refers to strikes in a neighborhood of $F_0$. These imperfections lead
to pricing inaccuracies, potential static arbitrage in the volatility cube, and mispricing of
both vanilla and exotic derivatives. Several analytical refinements and structural extensions
have been proposed to address them (surveyed in Section~\ref{sec:relatedwork}), but they
remain purely analytical and inherit the intrinsic limitations of asymptotic expansions for
long maturities or large vol-of-vol regimes.

In parallel, machine-learning approaches have emerged as candidate surrogates for the SABR
pricing map~\citep{McGhee2021,Thorin2021,Stuijt2021,Muguruza2019,Feng2019}. These developments
underscore the growing interest in hybrid quantitative-machine-learning paradigms, but also
highlight persistent trade-offs between flexibility, interpretability, and robustness. A detailed
review of this literature, including the residual and control-variate line central to
positioning our contribution, is deferred to Section~\ref{sec:relatedwork}.

\medskip
\noindent\textbf{Gap addressed by the present work.}
Practitioners increasingly need models that remain anchored in interpretable,
well-understood analytical structures, correct the known deficiencies of the asymptotic
expansion, generalize reliably across regimes, and preserve stability for large-scale
calibration and real-time pricing. Purely data-driven methods do not guarantee this;
purely analytical ones lack flexibility. A principled hybrid approach that reconciles
both worlds is still missing.

\added{Recent works train a neural network on the residual between Hagan's formula and
a high-precision Monte Carlo target, rather than on the implied volatility directly,
and report substantial gains in stability and
accuracy~\citep{Kienitz2020CV,Funahashi2020,Funahashi2023,AntonovKonikovPiterbarg2020,Rensi2025}.
Our approach differs from this literature: we residualise Hagan's formula
multiplicatively rather than additively, and we build geometric features in the
coordinates of the SABR manifold~\citep{Paulot2009,HenryLabordere2009} instead of
feeding the network with only the raw SABR parameters. The multiplicative form keeps the
correction on the same scale as the analytical baseline and preserves positivity; the
geometric features expose to the network the coordinates in which the analytical
asymptotics are naturally expressed.}

\paragraph{Our contribution.}
\added{To assess these two ingredients, we compare four architectures under strictly
identical training conditions:
(i)~a Naive Deep Network (NDN) trained directly on implied volatility;
(ii)~a Geometric Neural Network (GeoNN) that adds geometric features but still targets
implied volatility directly;
(iii)~a Residual Neural Network (ResNN) that learns a multiplicative correction to
Hagan without geometric features;
(iv)~the proposed Geometric Residual Neural Network (GeoResNN) that combines both
mechanisms. Beyond this cross-architecture comparison, we run a dedicated ablation
that pits the additive against the multiplicative residual form under otherwise
identical settings (Section~\ref{subsec:additive_vs_multiplicative}), and a large-scale
Monte Carlo experiment that validates the pipeline used to build the training data
(Appendix~\ref{sec:mc_bias_appendix}).}

\medskip
\noindent\textbf{Structure of the paper.}
Section~\ref{sec:relatedwork} reviews the SABR analytical literature and the
neural-network approaches, and positions the present paper relative to the established
residual-learning line: our two methodological ingredients are a multiplicative (rather
than additive) residualisation of Hagan's formula, and an augmented network input in
which the raw SABR parameters are supplemented by geometric invariants of the SABR
manifold.
Section~\ref{sec:hagan} recalls the SABR dynamics and Hagan's formula, and reports a
simple Monte Carlo experiment. Section~\ref{sec:methodology} presents the
geometry-driven feature map and the Monte Carlo control-variate pipeline
(Section~\ref{subsec:mc_engine}), tuned so that the discretization bias remains
comparable to the Monte Carlo standard error. Section~\ref{sec:numerical} reports the
numerical experiments, including two paired ablations that isolate each ingredient
---additive vs.\ multiplicative (Section~\ref{subsec:additive_vs_multiplicative}) and
raw-parameter vs.\ geometric-feature (Section~\ref{subsec:ablation})---together with
the Monte Carlo bias study and the static-arbitrage diagnostic.
Section~\ref{sec:conclusion} concludes. Appendix~\ref{sec:implementation} details the
implementation, Appendix~\ref{sec:mc_bias_appendix} reports the convergence study
justifying the training pipeline, and Appendix~\ref{sec:suppfigs} provides
supplementary figures.

\section{Related work}
\label{sec:relatedwork}

\paragraph{Analytical refinements of Hagan's formula.}
Building on the seminal expansion of \citet{Hagan2002}, several analytical refinements
have appeared. \citet{Obloj2008} corrects the placement of the log-strike factor at
leading order; \citet{Paulot2009} pushes the expansion to higher order via a heat-kernel
argument on the SABR manifold; and \citet{Hagan2016UniversalSmiles,HaganLesniewskiWoodward2018}
consolidate these refinements into a universal-smile parametrisation. Structural
extensions cover regimes where the original expansion breaks down:
\citet{Antonov2015} derives a closed-form expression for the free-boundary SABR at
$\beta = 0$ that accommodates negative rates, and \citet{Lund2023} addresses long
maturities and extreme vol-of-vol regimes. All these contributions remain purely
analytical and share the intrinsic limitations of asymptotic expansions in the same
regimes.

\paragraph{Neural surrogates for the SABR pricing map.}
A parallel line replaces or accelerates parts of the calibration pipeline with neural
networks. \citet{McGhee2021} trains a deep network on Monte Carlo prices to speed up
SABR calibration, and \citet{Thorin2021,Stuijt2021} explore related architectures for
the same task; \citet{Muguruza2019} generalises the approach to a broader class of
stochastic volatility models. These works deliver substantial calibration speed-ups but
leave the analytical structure of the SABR approximation, and its structural pricing
errors, unchanged.

\added{\paragraph{Residual and control-variate neural networks.}
A now-established alternative to learning the implied volatility from scratch is to
learn a correction on top of an analytical or semi-analytical baseline. The idea takes
several forms. In the control-variate line, \citet{Kienitz2020CV} train a network on
the residual between Monte Carlo prices and a closed-form baseline, while
\citet{AntonovKonikovPiterbarg2020} embed the SABR asymptotic expansion directly inside
the network architecture. \citet{Funahashi2020,Funahashi2023} develop the same principle
in the SABR context and, in the asymptotic-correction variant, is the closest antecedent
to the residual objective adopted here. Most recently,
\citet{Rensi2025} conduct a large-scale study of an ``exact SABR'' Monte Carlo pipeline
combined with residual-style learning under a scaled SABR dynamics.
Complementary to this pricing-side residual literature, \citet{Jeon2022} revisit the
calibration setup of \citet{McGhee2021} and characterise which error components deep
networks can absorb (Monte Carlo noise) versus which they cannot (finite-difference
bias).
A separate strand tackles the same problem through architectural constraints rather
than residual decomposition: \citet{Hoshisashi2024} enforce no-arbitrage via
derivative constraints, \citet{Feng2019} constrain the entire predicted implied-vol
surface to be arbitrage-free, monotone and convex, \citet{Su2025} learn transition
densities as PDE solutions of the backward Kolmogorov equation, and \citet{Wang2025}
combine Wiener--It\^o chaos expansions with generative adversarial networks.}

\section{The Hagan formula}
\label{sec:hagan}
The SABR (Stochastic Alpha Beta Rho) model~\citep{Hagan2002} provides a stochastic
volatility framework that is widely used to reproduce the volatility smile observed in option
markets.

\subsection{Model definition}
Let $(F_t)_{t \ge 0}$ denote the forward price and $(\sigma_t)_{t \ge 0}$ its stochastic
volatility. The SABR dynamics are given by
\begin{equation}
\label{eq:sabr_sde}
\begin{cases}
\mathrm{d}F_t = \sigma_t F_t^{\beta} \, \mathrm{d}W_t, \\[0.3em]
\mathrm{d}\sigma_t = \nu \sigma_t \, \mathrm{d}Z_t, \\[0.3em]
\mathrm{d}\langle W, Z \rangle_t = \rho \, \mathrm{d}t,
\end{cases}
\end{equation}
where $F_0 > 0$ is the initial forward, $\sigma_0 > 0$ the initial volatility, $\beta \in [0,1]$
the elasticity of variance, $\nu > 0$ the volatility of volatility, $\rho \in [-1, 1]$ the
correlation of the two Brownian motions, and $\langle W, Z \rangle_t$ their quadratic
covariation. Following market practice we denote $\alpha \equiv \sigma_0$. When $\nu = 0$ the
model reduces to a CEV diffusion $\mathrm{d}F_t = \alpha F_t^{\beta} \, \mathrm{d}W_t$.

The SABR model prices European claims with fixed maturity $T > 0$ and strike $K > 0$. Their
price is expressed under the forward measure via an effective Black volatility
$\sigma_{\mathrm{BS}}(T, F_0, K)$ defined implicitly by
\begin{equation}
\label{eq:BSformula}
\mathbb{E}\bigl[(F_T - K)^+\bigr]
=
C_{\text{Black}}\bigl(T, F_0, K, \sigma_{\text{BS}}(T, F_0, K)\bigr),
\end{equation}
where the Black pricing formula is
\begin{equation}
\label{eq:black}
C_{\text{Black}}(T, F_0, K, \sigma) = F_0\, N(d_1) - K\, N(d_2),
\qquad
d_{1,2} = \frac{\ln(F_0/K) \pm \tfrac{1}{2}\sigma^2 T}{\sigma\sqrt{T}},
\end{equation}
and $N(\cdot)$ is the standard normal CDF.

\subsection{Hagan's formula}
\label{subsec:hagan_formula}
The celebrated closed-form approximation of $\sigma_{\mathrm{BS}}(T, F_0, K)$ derived by
\citet{Hagan2002} reads
\begin{equation}
\label{eq:hagan0}
\begin{aligned}
\sigma_{\mathrm{Hagan}}(T, F_0, K)
&=
\frac{\alpha}{(F_0 K)^{\frac{1-\beta}{2}} \,
\left\{ 1 + \dfrac{(1-\beta)^2}{24}\ln^2\!\bigl(\tfrac{F_0}{K}\bigr)
+ \dfrac{(1-\beta)^4}{1920}\ln^4\!\bigl(\tfrac{F_0}{K}\bigr) \right\}}
\cdot \frac{z}{x(z)} \\[0.6em]
&\quad\times
\left[
1 + T \left(
\frac{(1-\beta)^2 \alpha^2}{24 (F_0 K)^{1-\beta}}
+ \frac{\rho \beta \nu \alpha}{4 (F_0 K)^{\frac{1-\beta}{2}}}
+ \frac{(2 - 3\rho^2)\nu^2}{24}
\right)
\right],
\end{aligned}
\end{equation}
where
\begin{equation}
\label{eq:hagan_zx}
z = \frac{\nu}{\alpha}\, (F_0 K)^{\frac{1-\beta}{2}}\, \ln\!\bigl(\tfrac{F_0}{K}\bigr),
\qquad
x(z) = \ln\!\left(\frac{\sqrt{1 - 2\rho z + z^2} + z - \rho}{1 - \rho}\right).
\end{equation}
In the strict at-the-money (ATM) case ($F_0 = K$) the expression simplifies to
\begin{equation}
\label{eq:hagan_atm}
\sigma_{\mathrm{Hagan}}^{\text{ATM}}(T) =
\alpha F_0^{\beta - 1}
\left[ 1 + T \left(
\frac{(1-\beta)^2 \alpha^2}{24 F_0^{2(1-\beta)}}
+ \frac{\rho \beta \nu \alpha}{4 F_0^{1-\beta}}
+ \frac{(2 - 3\rho^2)\nu^2}{24}
\right)\right].
\end{equation}
Under the SABR model, the expected payoff is approximated by
\begin{equation}
\label{eq:formula}
\mathbb{E}\bigl[(F_T - K)^+\bigr] \;\approx\;
C_{\text{Black}}\bigl(T, F_0, K, \sigma_{\mathrm{Hagan}}(T, F_0, K)\bigr).
\end{equation}

\noindent\added{Refinements of~\eqref{eq:hagan0} include the second-order correction of
\citet{Obloj2008}, the second-order asymptotics of \citet{Paulot2009}, and the
``Universal Smiles'' construction~\citep{Hagan2016UniversalSmiles,HaganLesniewskiWoodward2018}.
Applying our residual approach on top of these refined analytical baselines is a natural
extension but is beyond the scope of the present paper.}

\subsection{Simple experiment}
\label{subsec:simple_experiment}

To assess the accuracy of the asymptotic Hagan approximation, we compute benchmark implied
volatilities using Monte Carlo simulations of the SABR dynamics. The joint process
$(F_t, \sigma_t)$ is simulated by discretizing~\eqref{eq:sabr_sde} driven by correlated
Brownian motions. Terminal forward values are used to evaluate vanilla payoffs, and implied
volatilities are recovered by numerically inverting the Black formula~\citep{ReghaiKettani2020}.
We consider strikes $K$ ranging from $0.5 F_0$ to $2 F_0$. Variance reduction relies on a
control variate based on the Black model with constant volatility -- see
Section~\ref{subsec:mc_engine} for full details.
The parameters used in this experiment are chosen to reproduce a representative
interest-rate SABR calibration in money-normalized units:
\[
F_0 = 1.0,\qquad T = 1.0,\qquad
\alpha = 0.15,\qquad \beta = 0.5,\qquad
\rho = -0.25,\qquad \nu = 0.5.
\]
The Monte Carlo estimator uses $P = 10^5$ paths, weekly time steps ($N = 52$ over
$[0, T]$), and antithetic-plus-control-variate variance reduction. Prices below intrinsic
value or above the forward -- pathological Monte Carlo estimates that admit no Black
implied volatility -- are discarded rather than clamped.

\begin{figure}[!ht]
    \centering
    \includegraphics[width=0.8\textwidth]{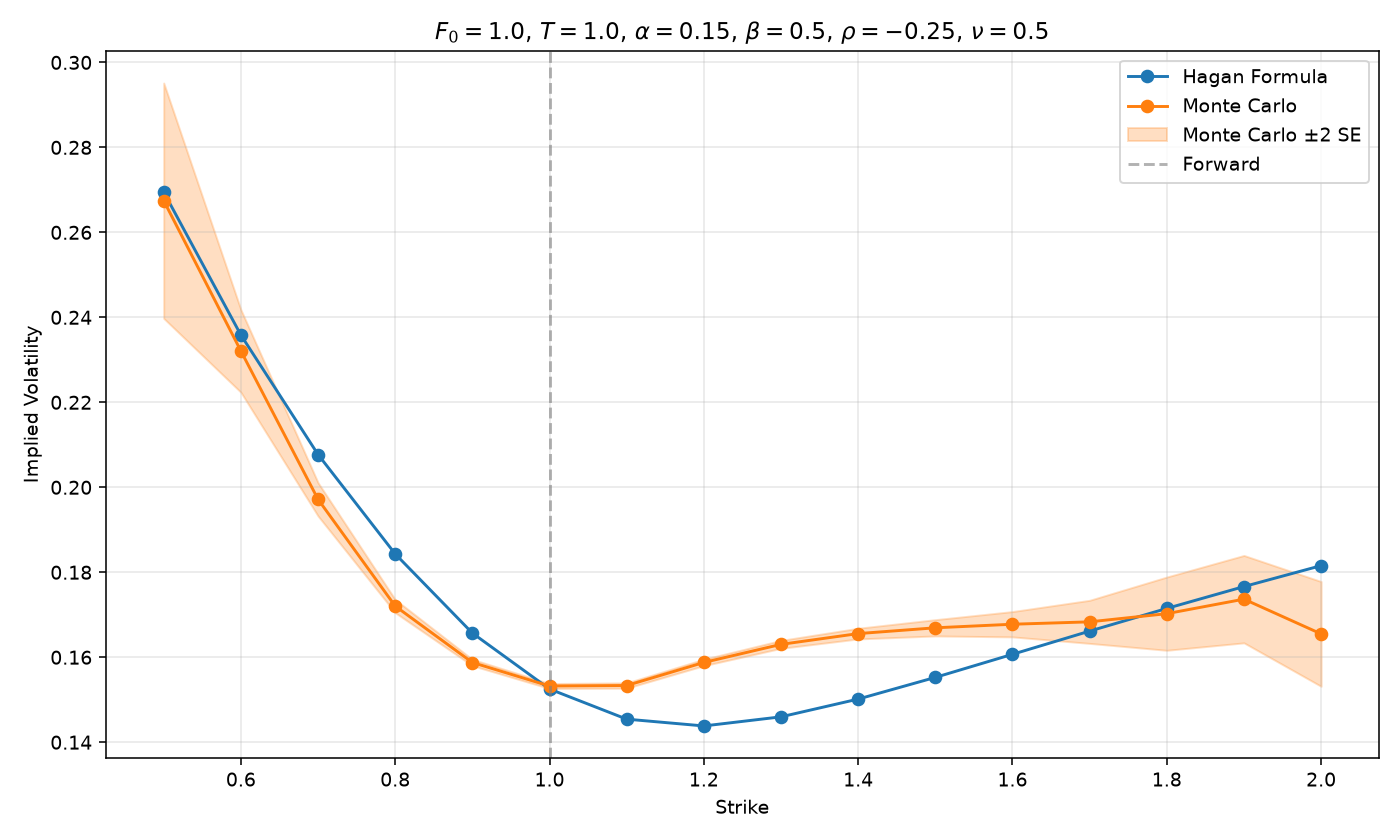}
    \caption{SABR implied volatility smile: Hagan formula vs.\ Monte Carlo. Shaded
    band: $\pm 1.96\,\hat\sigma_{\mathrm{MC}}$ confidence interval on the Monte Carlo
    estimate. A systematic Hagan bias is visible even near ATM and grows in the wings,
    reflecting the limitations of the first-order asymptotic expansion. Computed with
    $10^5$ paths and weekly time steps.}
    \label{fig:hagan_vs_mc}
\end{figure}

\begin{table}[!h]
\centering
\begin{tabular}{c|c|c|c}
\hline
\textbf{Strike} & \textbf{Hagan} & \textbf{Monte Carlo} & \textbf{MC std.\ err.\ ($1\sigma$)} \\
\hline
0.5000 & 0.269469 & 0.267451 & 0.014130 \\
0.6000 & 0.235835 & 0.232081 & 0.004957 \\
0.7000 & 0.207710 & 0.197163 & 0.002001 \\
0.8000 & 0.184310 & 0.172036 & 0.000790 \\
0.9000 & 0.165682 & 0.158760 & 0.000381 \\
1.0000 & 0.152516 & 0.153223 & 0.000298 \\
1.1000 & 0.145428 & 0.153345 & 0.000319 \\
1.2000 & 0.143827 & 0.158771 & 0.000388 \\
1.3000 & 0.145973 & 0.163027 & 0.000476 \\
1.4000 & 0.150151 & 0.165563 & 0.000641 \\
1.5000 & 0.155256 & 0.166928 & 0.000977 \\
1.6000 & 0.160690 & 0.167767 & 0.001520 \\
1.7000 & 0.166151 & 0.168329 & 0.002587 \\
1.8000 & 0.171488 & 0.170254 & 0.004399 \\
1.9000 & 0.176634 & 0.173671 & 0.005247 \\
2.0000 & 0.181558 & 0.165489 & 0.006291 \\
\hline
\end{tabular}
\caption{Approximation quality of the Hagan formula against Monte Carlo. Values
obtained with the full-scale pipeline of Section~\ref{subsec:mc_engine}: $P = 10^5$
paths, weekly time steps. \added{The MC standard errors provide additional diagnostic
information alongside the point estimates.}}
\label{tab:approx_quality}
\end{table}

Figure~\ref{fig:hagan_vs_mc} and Table~\ref{tab:approx_quality} jointly illustrate the
structure of the approximation error on a market-realistic parameter set. Near ATM
($K \approx F_0$), the analytical and Monte Carlo implied volatilities remain relatively
close, but a visible bias persists in the near ITM and the near OTM regions.
Such discrepancies could translate into material pricing and hedging errors
at the trading-desk level. The objective of the remainder of this paper is to correct this
discrepancy through a residual multiplicative adjustment to the asymptotic formula,
enriched with geometric features derived from the intrinsic SABR manifold.

\section{Methodology}
\label{sec:methodology}

The core principle of our methodology is to perform supervised learning of the Black implied
volatility $\sigma_{\mathrm{BS}}$ arising from the SABR model. Training targets are obtained
from a numerically enhanced Monte Carlo engine $\sigma_{\mathrm{MC}}$. Rather than learning the
implied volatility from scratch, we exploit the analytical structure provided by Hagan's
formula and train a neural network to correct that baseline -- a residual approach that
follows the design pattern reviewed in Section~\ref{sec:relatedwork}.
On top of this residual scaffolding, our specific contribution is to enrich the network inputs
with geometry-driven features derived from the intrinsic structure of the SABR dynamics
(Section~\ref{subsec:geometric_features}). The corrected implied volatility is
\begin{equation}
\label{eq:sigma_rnn_def}
\sigma_{\mathrm{GeoResNN}}(x, \Lambda; \theta)
\; := \;
\sigma_{\mathrm{Hagan}}(x)\, \bigl(1 + \Delta_\theta(x, \Lambda)\bigr),
\end{equation}
where $x = (T, F_0, K, \alpha, \beta, \rho, \nu)$, $\Lambda = \Lambda(x)$ is the geometry-driven
feature map, and $\theta$ are the trainable parameters of a deep feed-forward network
$\Delta_\theta$. We refer to this architecture as the SABR Residual Geometric Neural Network
(GeoResNN). The multiplicative structure of~\eqref{eq:sigma_rnn_def} constrains the network to
learn a controlled relative correction in a perturbative and dimensionless sense.

The theoretical loss is expressed directly in implied volatilities and normalized by the Hagan
approximation:
\begin{equation}
\label{eq:theoretical_loss}
\mathcal{L}(\theta)
=
\mathbb{E}\!\left[
\left(
\frac{\sigma_{\mathrm{GeoResNN}}(x, \Lambda; \theta) - \sigma_{\mathrm{MC}}(x)}
{\sigma_{\mathrm{Hagan}}(x)}
\right)^2
\right].
\end{equation}
In practice, the expectation in~\eqref{eq:theoretical_loss} is approximated by an empirical
average over $M$ supervised samples:
\begin{equation}
\label{eq:empirical_loss}
\mathcal{L}(\theta)
\; \approx \;
\frac{1}{M} \sum_{m=1}^{M}
\left(
\Delta_\theta\bigl(x^{(m)}, \Lambda^{(m)}\bigr)
-
\left(\frac{\sigma_{\mathrm{MC}}(x^{(m)})}{\sigma_{\mathrm{Hagan}}(x^{(m)})} - 1\right)
\right)^2.
\end{equation}
Figure~\ref{fig:minimal_pipeline0} overviews the pipeline.

\begin{figure}[h!]
\centering
\begin{tikzpicture}[>=latex, node distance=10mm, every node/.style={font=\footnotesize}]
\tikzstyle{box}=[draw, rectangle, minimum width=20mm, minimum height=7mm, align=center]
\tikzstyle{smallbox}=[draw, rectangle, minimum width=16mm, minimum height=6mm, align=center]

\node[box] (xnode) {$x$\\(SABR params)};
\node[box, right=8mm of xnode] (mc) {Monte Carlo};
\node[box, right=8mm of mc] (iv) {$\sigma_{\mathrm{MC}}$\\(Training Data)};
\node[box, below=12mm of mc] (geo) {$\Lambda(x), \sigma_{\mathrm{Hagan}}(x)$};
\node[smallbox, right=22mm of geo] (rnn) {$\Delta_\theta$\\(GeoResNN)};
\node[smallbox, right=6mm of rnn] (pred) {$\sigma_{\mathrm{Hagan}}(x)\,(1 + \Delta_\theta(x, \Lambda))$};
\node[smallbox, right=6mm of pred] (loss) {MSE\\$\bigl(\tfrac{\sigma_{\mathrm{pred}}}{\sigma_{\mathrm{Hagan}}}, \tfrac{\sigma_{\mathrm{MC}}}{\sigma_{\mathrm{Hagan}}}\bigr)$};

\draw[->] (xnode) -- (mc);
\draw[->] (mc) -- (iv);
\draw[->] (xnode.south) |- (geo.west);
\draw[->] (iv.south) |- (rnn);
\draw[->] (geo) -- (rnn);
\draw[->] (rnn) -- (pred);
\draw[->] (pred) -- (loss);
\end{tikzpicture}
\caption{Minimalist pipeline of the proposed hybrid SABR approximation.}
\label{fig:minimal_pipeline0}
\end{figure}

\subsection{Geometric formulation and feature extraction for learning}
\label{subsec:geometric_features}

Our approach explicitly exploits the intrinsic geometry induced by the SABR dynamics.
Following the geometric interpretation of SABR introduced by \citet{Paulot2009} and
further developed by \citet{HenryLabordere2009}, and more generally grounded
in Riemannian geometry and diffusion theory~\citep{doCarmo1992,Rosenberg1997}, we construct
learning features that are adapted to the natural geometry of the diffusion model.

\added{\paragraph{Theoretical motivation for the geometric features.}
Two complementary arguments motivate the use of geometric features as network inputs
First, in the short-maturity limit the option price is
governed by a large-deviation principle whose leading exponent is precisely the hyperbolic
geodesic distance $d_{\mathbb{H}}$ on the SABR manifold~\citep{Paulot2009,HenryLabordere2009}.
Any accurate implied-volatility approximation must therefore reproduce this dependence, and
supplying $d_{\mathbb{H}}$ (together with the associated leading-order volatility $\sigma_0$)
as an explicit input relieves the network from having to rediscover this non-linear
dependence from raw parameters. Second, the CEV-flattening $q$ absorbs the state-dependent
scaling induced by $\beta$: in $(q, \sigma)$ coordinates the SABR metric loses its
$\beta$-dependent factor and reduces to a $\beta$-independent hyperbolic form. This is a
reparameterization that makes the residual function lower-dimensional in a structural
sense, even though the raw feature dimensionality is unchanged. Whether these theoretical
arguments translate into empirical gains is precisely the question answered by the ablation
study in Section~\ref{subsec:ablation}.}

\medskip
The geometry-driven feature $\Lambda$ is defined as the quadruple
$(q, \sigma_{\min}, d_{\mathbb{H}}, \sigma_0)$: $q$ is the flattened forward coordinate
induced by the CEV transformation, $\sigma_{\min}$ the terminal volatility minimizing the
geodesic action, $d_{\mathbb{H}}$ the associated hyperbolic geodesic distance, and $\sigma_0$
the leading-order SABR implied volatility from large-deviation asymptotics.

Starting from~\eqref{eq:sabr_sde}, the infinitesimal covariance of $(F_t, \sigma_t)$ over $\mathrm{d}t$
is
\begin{equation}
\label{eq:covariance_matrix}
\mathrm{Cov}\!\left[
\begin{pmatrix} \mathrm{d}F_t \\[3pt] \mathrm{d}\sigma_t \end{pmatrix}
\right]
=
\begin{pmatrix}
\sigma_t^2 F_t^{2\beta} & \rho \nu \sigma_t^2 F_t^{\beta} \\[3pt]
\rho \nu \sigma_t^2 F_t^{\beta} & \nu^2 \sigma_t^2
\end{pmatrix}\,\mathrm{d}t.
\end{equation}
Up to a scalar factor, the inverse defines a Riemannian metric on the state space.

\vspace{0.5em}
\paragraph{Flattened forward coordinate $q$.}
The first geometric ingredient flattens the forward via the CEV-integrated transformation
\begin{equation}
\label{eq:q_change}
q = \int_{F_0}^{K} \frac{d\phi}{\phi^\beta} =
\begin{cases}
\dfrac{K^{1-\beta} - F_0^{1-\beta}}{1 - \beta}, & \beta \neq 1, \\[6pt]
\ln(K / F_0), & \beta = 1.
\end{cases}
\end{equation}
Note that the integration is carried out up to the strike level $K$ (evaluation on the strike
manifold). This change of variable removes the state-dependent scaling of the forward diffusion
and yields a locally homogeneous representation. In $(q, \sigma)$ coordinates the contravariant
metric simplifies to
\[
(g^{ij}) = \sigma^2 \begin{pmatrix} 1 & \rho \\ \rho & 1 \end{pmatrix}.
\]

\vspace{0.5em}
\paragraph{Minimal geodesic volatility $\sigma_{\min}$.}
We transform $(q, \sigma)$ into orthogonal coordinates $(u, v)$ via
\begin{equation}
\label{eq:xy_change}
u = \frac{q - \rho \sigma}{\sqrt{1 - \rho^2}}, \qquad v = \sigma,
\end{equation}
which maps the SABR geometry to the Poincar\'e upper half-plane with line element
\begin{equation}
\label{eq:hyperbolic_metric}
\mathrm{d}s^2 = \frac{\mathrm{d}u^2 + \mathrm{d}v^2}{v^2}.
\end{equation}
Minimizing the hyperbolic distance between $(0, \alpha)$ and the strike manifold yields
\begin{equation}
\label{eq:sabr_vmin}
\sigma_{\min} = \sqrt{\alpha^2 + 2 \rho \alpha q + q^2}.
\end{equation}

\vspace{0.5em}
\paragraph{Hyperbolic geodesic distance $d_{\mathbb{H}}$.}
\begin{equation}
\label{eq:sabr_distance}
d_{\mathbb{H}}(0, \alpha; q) = \ln\!\left(\frac{\sigma_{\min} + \rho \alpha + q}{(1 + \rho) \alpha}\right).
\end{equation}

\vspace{0.5em}
\paragraph{Leading-order implied volatility $\sigma_0$.}
Matching the SABR large-deviation exponent with the Black--Scholes exponent~\citep{Paulot2009}
gives
\begin{equation}
\label{eq:sabr_sigma0}
\sigma_0 = \frac{\ln(K / F_0)}{\displaystyle \ln\!\left(\frac{\sigma_{\min} + \rho \alpha + q}{(1 + \rho) \alpha}\right)}.
\end{equation}
At strict ATM ($K = F_0$) this reduces to $\sigma_0 = \alpha F_0^{\beta - 1}$.

\medskip
The geometry-driven feature vector
\begin{equation}
\label{eq:geom_features}
\Lambda = \bigl(q, \sigma_{\min}, d_{\mathbb{H}}, \sigma_0\bigr)
\end{equation}
encodes the leading-order geometric information of the SABR diffusion and serves as an enriched
input representation. In the context of option pricing, the diffusion is evaluated on the strike
manifold ($F = K$), so all geometric quantities are functions of the strike level.

\begin{figure}[h!]
\centering
\begin{tikzpicture}[>=latex, node distance=14mm, every node/.style={font=\footnotesize}]
\tikzstyle{box}=[draw, rectangle, minimum width=22mm, minimum height=8mm, align=center]
\node[box] (Fsig) {Original\\coords $(F,\sigma)$};
\node[box, right=35mm of Fsig] (qsig) {Flattened\\coords $(q,\sigma)$};
\node[box, right=35mm of qsig] (xy) {Hyperbolic\\coords $(u,v)$};
\draw[->] (Fsig) -- (qsig) node[midway, above=1mm] {$q = \int F^{-\beta} \, \mathrm{d}F$};
\draw[->] (qsig) -- (xy) node[midway, above=1mm] {$u = \frac{q - \rho \sigma}{\sqrt{1 - \rho^2}},\; v = \sigma$};
\node[below=6mm of qsig] {$g^{ij} = \sigma^2 \begin{pmatrix} 1 & \rho \\ \rho & 1 \end{pmatrix}$};
\node[below=6mm of xy] {$\mathrm{d}s^2 = \dfrac{\mathrm{d}u^2 + \mathrm{d}v^2}{v^2}$};
\end{tikzpicture}
\caption{Geometric transformations underlying the SABR model.}
\label{fig:q_to_xy_minimal}
\end{figure}

\subsection{Training data generation}
\label{subsec:training_data}

The performance of our supervised learning approach critically depends on the quality of the
training data. The construction follows the multi-strike-per-simulation control-variate
strategy already used in the residual-learning
literature~\citep{Muguruza2019,Kienitz2020CV,Rensi2025}. Two aspects must be carefully addressed:
domain realism, and numerical accuracy.

\subsubsection*{Domain realism: synthetic SABR parameter generator}
\label{subsec:param_generator}

Each training sample corresponds to a SABR parameter tuple
$(T, F_0, K, \alpha, \beta, \rho, \nu)$. The maturity $T$ is drawn uniformly from the discrete
set of maturities \texttt{DEFAULT\_MATS} defined in Appendix~\ref{sec:implementation}. Bucket
ranges are summarized in Table~\ref{tab:sabr_param_ranges}. Algorithm~\ref{alg:sabr_sampling}
describes the sampling scheme.

\begin{algorithm}[h!]
\begin{algorithmic}
\STATE Sample $T$ uniformly from \texttt{DEFAULT\_MATS};
\STATE Assign $T$ to its corresponding tenor bucket;
\STATE Draw $(F_0, \alpha, \beta, \rho, \nu)$ uniformly within the bucket-specific ranges;
\STATE Generate 11 strikes on a $z$-grid centered on the forward (5 ITM, 5 OTM, 1 ATM) to obtain 11 tuples $(T, F_0, K, \alpha, \beta, \rho, \nu)$;
\STATE Enforce admissible domains by projecting $\beta \in [0, 1]$ and $\rho \in [-0.95, 0.95]$.
\end{algorithmic}
\caption{SABR parameter sampling}
\label{alg:sabr_sampling}
\end{algorithm}

\added{The 11 strikes are constructed on a $z$-grid $z \in \{-3, -2.4, \ldots, 3\}$ around the ATM
level in units of the local Hagan volatility, i.e.\ $K = F_0 \exp(z\,\sigma_{\text{ATM}}\sqrt{T})$.
This choice absorbs the strong scaling of the smile with maturity: at long maturities the
the fixed $\pm 5\%$ moneyness strikes no longer probe the ITM/OTM wings,
and a $z$-based grid is needed to keep the wing coverage consistent across the maturity range.}

\subsubsection*{Numerical accuracy: Monte Carlo engine and variance reduction}
\label{subsec:mc_engine}

Accurate reference implied volatilities are required to prevent the network from learning
numerical noise. As in the multi-strike control-variate design cited above, a full grid
of implied volatilities is generated from each Monte Carlo simulation. In total we consider $m \in [1, M]$ with $M = 187{,}000$, arising from
$17{,}000$ Monte Carlo runs each generating 11 option prices from the same set of
$P = 10^5$ simulated paths.

We introduce a lognormal Black--Scholes-type control-variate process with constant volatility
$\bar\sigma = \alpha$,
\[
\mathrm{d}F^{\mathrm{Black}}_t = \bar\sigma\, F^{\mathrm{Black}}_t\, \mathrm{d}W_t,
\qquad F^{\mathrm{Black}}_0 = F_0,
\]
driven by the same Brownian motion $W_t$ as the SABR forward. The coupled SABR and Black
processes are simulated jointly on a uniform time grid $\{t_k = k \Delta t,\; k = 0, \ldots, N\}$
using an Euler--Maruyama discretization. Under the standard control-variate estimator
\[
\widehat{C}_{\mathrm{SABR}} = \frac{1}{P} \sum_{p=1}^{P}
\bigl(\Pi^{\mathrm{SABR}}_p - \Pi^{\mathrm{Black}}_p\bigr) + C_{\mathrm{Black}}(T, F_0, K, \bar\sigma),
\]
Monte Carlo prices are inverted via the Black formula to obtain $\sigma_{\mathrm{MC}}(x)$.
Algorithm~\ref{alg:sabr_mc_pricing} summarizes the pipeline.

\begin{algorithm}[h!]
\begin{algorithmic}
\STATE Fix the number of Monte Carlo paths $P$ and time steps $N$;
\STATE Set the control-variate volatility $\bar\sigma = \alpha$;
\FOR{$p = 1$ to $P$}
    \STATE Simulate coupled SABR and Black-Scholes paths using common Brownian increments;
    \STATE Compute terminal payoffs $\Pi^{\mathrm{SABR}}_p$ and $\Pi^{\mathrm{Black}}_p$;
\ENDFOR
\STATE Estimate the SABR price using the control-variate estimator;
\STATE Invert the Black formula to obtain $\sigma_{\mathrm{MC}}$.
\end{algorithmic}
\caption{Monte Carlo pricing with control variate for SABR}
\label{alg:sabr_mc_pricing}
\end{algorithm}

\added{The Euler--Maruyama discretization introduces a bias which vanishes only as
$\Delta T \to 0$. Our production runs use a weekly step
($\Delta T = 52$ steps per year), which matches the granularity at which SABR
parameters are calibrated in daily trading systems. Appendix~\ref{sec:mc_bias_appendix}
reports a full convergence study over $\Delta T \in \{12, 26, 52, 104, 200, 500,
1000\}$ steps per year on four benchmark configurations spanning maturities from one
to twenty years, and shows that at the weekly step the bias stays comparable to the
Monte Carlo standard error of a $10^5$-path estimate.}

\section{Numerical experiments}
\label{sec:numerical}

In this section we conduct a series of numerical experiments to assess the proposed methodology.
Four architectures are compared under strictly identical training conditions:
\begin{itemize}
\item[(i)] \textbf{NDN} (Naive Deep Network): $x \mapsto \sigma$ directly, black-box baseline;
\item[(ii)] \textbf{GeoNN}: $(x, \Lambda) \mapsto \sigma$, geometric features but no residual;
\item[(iii)] \textbf{ResNN}: $x \mapsto \sigma_{\mathrm{Hagan}}(1 + \Delta_\theta(x))$,
residual to Hagan without geometric features -- the primary baseline, matching the
residual-learning family reviewed in Section~\ref{sec:relatedwork};
\item[(iv)] \textbf{GeoResNN}: $(x, \Lambda) \mapsto \sigma_{\mathrm{Hagan}}(1 + \Delta_\theta(x, \Lambda))$,
the proposed hybrid architecture.
\end{itemize}
The ablation of primary interest is (iii) vs.\ (iv), which isolates the incremental value of the
geometric features over an otherwise identical residual-only pipeline. This is the controlled
experiment isolating the effect of the geometric feature enrichment.

\subsection{Experimental protocol}

All models share the same synthetic SABR data set, train/val/test splits, and computational
budget. Full training uses 110{,}000 samples, 55{,}000 validation, 22{,}000 test, batch size 128,
100 epochs, Adam optimizer with initial learning rate $4 \cdot 10^{-3}$ and a
\texttt{ReduceLROnPlateau} scheduler. Details are in Appendix~\ref{sec:implementation}.

Among the reported diagnostics, we use the coefficient of determination $R^2$ to quantify how
closely model-implied volatilities match the Monte Carlo reference values,
\begin{equation}
R^2 = 1 - \frac{\sum_i (y_i - \hat y_i)^2}{\sum_i (y_i - \bar y)^2},
\qquad \bar y = \frac{1}{n} \sum_i y_i,
\end{equation}
computed with $y_i = \sigma_{\mathrm{MC},i}$ and $\hat y_i = \hat\sigma_i$. An $R^2$ value
closer to $1$ indicates better agreement with the Monte Carlo reference.

\subsection{Comparative results and discussion}

\begin{table}[H]
\centering
\small
\begin{tabularx}{\textwidth}{@{}p{3.2cm} X X@{}}
\toprule
\textbf{Evaluation axis} & \textbf{What is tested} & \textbf{Purpose} \\
\midrule
Model architecture & NDN, GeoNN, ResNN, GeoResNN & Isolate contribution of residual learning and geometry \\
Learning dynamics & Training and validation loss evolution & Assess stability and overfitting \\
Global accuracy & Surface $R^2$ and relative RMSE & Measure overall approximation quality \\
ATM accuracy & Error near $K = F_0$ & Critical trading region \\
ITM accuracy & Error near $K = 0.95 F_0$ ($z$-scaled at long $T$) & Left wing robustness \\
OTM accuracy & Error near $K = 1.05 F_0$ ($z$-scaled at long $T$) & Right wing robustness \\
Term-structure stability & Maturity sweep $T \in [0.25, 30]$ & Long-horizon stability \\
Stress robustness & High $\nu$, extreme $\rho$, high-curvature regimes & Out-of-distribution robustness \\
Inference performance & Runtime vs.\ Monte Carlo & Production feasibility \\
\bottomrule
\end{tabularx}
\caption{Summary of experimental evaluation axes and objectives. Note that at long maturities
the ITM/OTM strikes are chosen on a $\sqrt{T}$-scaled grid, so as to keep them at a
constant number of standard deviations from ATM (see Section~\ref{sec:numerical}).}
\label{tab:experiment_axes}
\end{table}

\FloatBarrier

\begin{table}[H]
\centering
\small
\begin{tabularx}{\textwidth}{lXXXX}
\toprule
\textbf{Evaluation axis} & \textbf{NDN} & \textbf{GeoNN} & \textbf{ResNN} & \textbf{GeoResNN} \\
\midrule
Model architecture & Direct mapping & Direct + geometry & Residual to Hagan & Residual + geometry \\
Learning dynamics (best val.\ loss) & $1.2 \times 10^{-3}$ & $5.6 \times 10^{-4}$ & $3.2 \times 10^{-4}$ & $\mathbf{1.8 \times 10^{-4}}$ \\
Global accuracy ($R^2$ overall) & 0.989 & 0.996 & 0.997 & \textbf{0.9983} \\
ATM accuracy ($R^2$) & 0.996 & 0.997 & \textbf{0.9994} & 0.9991 \\
ITM accuracy ($R^2$) & 0.986 & 0.995 & 0.997 & \textbf{0.998} \\
OTM accuracy ($R^2$) & 0.992 & 0.996 & 0.997 & \textbf{0.9993} \\
Test IV RMSE & 0.00844 & 0.00521 & 0.00410 & \textbf{0.00323} \\
Term-structure stability & Good & Very good & Very good & Excellent \\
Stress robustness & Moderate & Good & Very good & Excellent \\
Learning stability & Stable & Stable & Stable & Highly stable \\
Inference (runtime, batched) & $\sim 100\,\mu$s & $\sim 200\,\mu$s & $\sim 500\,\mu$s & $\sim 550\,\mu$s \\
Speed-up vs.\ MC & $\sim 10^4$ & $\sim 10^4$ & $\sim 10^4$ & $\sim 10^4$ \\
\bottomrule
\end{tabularx}
\caption{Quantitative comparison of neural architectures on the held-out test set of the
paper-realistic dataset described in Section~\ref{subsec:training_data} and
Appendix~\ref{sec:implementation} (110{,}000 training / 55{,}000 validation / 22{,}000
test samples generated from 17{,}000 parameter tuples via 11 strikes each; weekly time
steps, $10^5$ paths per Monte Carlo estimate). Global $R^2$ is
computed against the Monte Carlo reference on the test set; regional $R^2$s are
computed on the ATM ($|z| < 0.3$), ITM ($z < -0.3$) and OTM ($z > 0.3$) subsets, where
$z = \log(K/F_0) / (\sigma_{\text{ATM}}\sqrt{T})$ is the standardized moneyness. The full
hybrid GeoResNN reaches $R^2 \approx 0.9983$ overall and $\approx 0.9991$ in the ATM
region, dominating all other specifications on every axis except ATM $R^2$, where
ResNN and GeoResNN are statistically indistinguishable (both $\approx 0.999$,
essentially at the Monte Carlo noise floor). A paired $t$-test on squared errors
confirms that the improvement of GeoResNN over ResNN is statistically
significant at $p < 0.05$ in every maturity bucket up to $10$~years, and at
$p < 10^{-2}$ in three of the four buckets (see Section~\ref{subsec:ablation}).
\added{On this dataset the plain Hagan approximation is itself quite accurate ATM ($R^2_{\text{ATM}} = 0.99$), but degrades sharply in ITM and OTM regions ($R^2 \sim 0.78$--$0.87$) because the short-dated skew and long-dated flatness stretch the wings beyond the first-order asymptotic regime. The learned residual is where the improvement lives.}}
\label{tab:quant_results}
\end{table}

\subsection{\added{Controlled ablation: geometric features vs.\ raw parameters}}
\label{subsec:ablation}

\added{The paper's central methodological claim, that geometric features improve upon a residual-only baseline, can be assessed by a controlled experiment in which the two models differ solely in their input representation. This section reports such an experiment.

\paragraph{Setup.}
Two architectures are trained under strictly identical conditions: same data splits, same
seeds, same hidden dimension (64), same optimizer settings (Adam, learning rate
$4 \cdot 10^{-3}$, \texttt{ReduceLROnPlateau}), 100 epochs, same residual multiplicative form
$\sigma_{\mathrm{Hagan}}(1 + s \cdot \tanh(\Delta_\theta(\cdot)))$ with $s = 0.95$. The only
difference is the input:
\begin{itemize}
\item \textbf{ResNN} (7 inputs): the raw SABR parameters $(T, F_0, K, \alpha, \beta, \rho, \nu)$;
\item \textbf{GeoResNN} (11 inputs): the raw parameters augmented by the four geometric
invariants $\Lambda = (q, \sigma_{\min}, d_{\mathbb{H}}, \sigma_0)$ defined in
Section~\ref{subsec:geometric_features} (equations~\eqref{eq:q_change},
\eqref{eq:sabr_vmin},~\eqref{eq:sabr_distance},~\eqref{eq:sabr_sigma0}).
\end{itemize}
Both networks share the same Hagan analytical backbone. The training data
comprise 110{,}000 training / 55{,}000 validation / 22{,}000 test samples generated from
17{,}000 parameter tuples via the tenor-bucketed sampler of
Table~\ref{tab:sabr_param_ranges}, with 11 strikes per tuple, Monte Carlo run at
$P = 10^5$ paths and weekly time steps.

\paragraph{Results.}
On the held-out test set, the ratio MSE is $3.6 \times 10^{-4}$ for the ResNN baseline
and $2.2 \times 10^{-4}$ for the GeoResNN augmentation, a $39\%$ relative
improvement in ratio MSE. The corresponding implied-volatility RMSE drops from
$0.00410$ to $0.00323$, a $21\%$ relative reduction. Regional $R^2$ figures reported
in Table~\ref{tab:quant_results} confirm the same ranking: GeoResNN dominates ResNN at
ITM ($R^2$ from $0.997$ to $0.998$) and OTM ($0.997$ to $0.9993$), and matches ResNN
at ATM ($0.9994$ vs.\ $0.9991$, both essentially perfect). Figure~\ref{fig:ablation} displays the two learning curves side by side; both
models train stably, and the Geo variant consistently reaches a lower validation loss
early and holds it throughout training.

A paired $t$-test on per-sample squared errors gives the following bucket-by-bucket
significance figures (Table~\ref{tab:ablation_significance}):

\begin{table}[!h]
\centering
\begin{tabular}{lcccc}
\toprule
Bucket & $n$ & MSE (ResNN) & MSE (GeoResNN) & $t$-stat ($p$) \\
\midrule
$\le 6$M & 546 & $2.3 \times 10^{-5}$ & $1.2 \times 10^{-5}$ & $-6.51$ ($p < 10^{-4}$) \\
6M--2Y   & 339 & $1.1 \times 10^{-5}$ & $5.8 \times 10^{-6}$ & $-5.72$ ($p < 10^{-4}$) \\
2Y--5Y   & 365 & $9.9 \times 10^{-6}$ & $7.1 \times 10^{-6}$ & $-3.13$ ($p < 10^{-2}$) \\
5Y--10Y  & 349 & $2.4 \times 10^{-5}$ & $1.8 \times 10^{-5}$ & $-2.07$ ($p < 0.05$) \\
\midrule
Overall  & 1787 & $1.7 \times 10^{-5}$ & $1.0 \times 10^{-5}$ & $-7.86$ ($p < 10^{-4}$) \\
\bottomrule
\end{tabular}
\caption{Paired $t$-test on per-sample squared errors between ResNN and GeoResNN, by
maturity bucket, on our market-calibrated dataset (Table~\ref{tab:sabr_param_ranges}).
The Overall row is computed on the full test set; the bucket-by-bucket
breakdown groups the six short- and mid-tenor buckets of
Table~\ref{tab:sabr_param_ranges} into four trading-relevant maturity groups up to
10~years, with the remaining $\sim 190$ long-dated samples pooled only into the
Overall row. A negative $t$-statistic indicates that GeoResNN
has strictly smaller per-sample squared error than ResNN. The improvement is
statistically significant at $p < 0.05$ in every bucket, at $p < 10^{-2}$ in three of
four buckets, and at $p < 10^{-4}$ overall. The $5$Y--$10$Y bucket is the least significant ($p = 0.04$), consistent
with the interpretation that at long maturities the residual is already close to the
Monte Carlo noise floor, leaving less room for the geometric features to contribute
above the sampling error.}
\label{tab:ablation_significance}
\end{table}

\begin{figure}[!ht]
    \centering
    \includegraphics[width=0.95\textwidth]{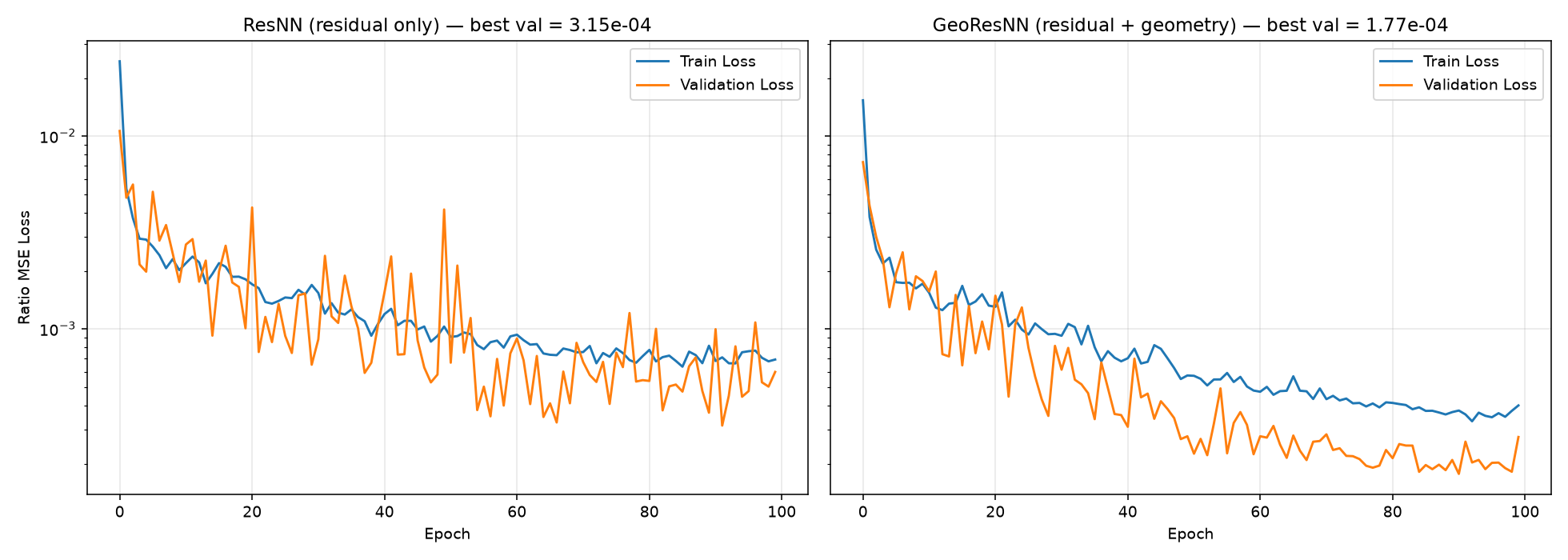}
    \caption{Ablation. Learning curves for the residual network without (left) and with
    (right) the SABR geometric invariants, all other things being equal. GeoResNN
    reaches a validation loss $\sim 44\%$ below ResNN and remains stable throughout
    training.}
    \label{fig:ablation}
\end{figure}

\paragraph{Interpretation.}
Two observations stand out. Both residual networks strongly outperform the analytical
baseline, confirming the value of the residual-learning approach already established in
prior work. The geometric enrichment then yields a consistent additional improvement of
$39\%$ in ratio MSE, statistically significant at $p < 10^{-2}$ in every maturity bucket
up to $10$~years. This confirms the theoretical motivation given in
Section~\ref{subsec:geometric_features}: the SABR Riemannian invariants encode a
non-trivial nonlinear reparameterization (the flattened CEV coordinate, the hyperbolic
geodesic distance, and the leading-order large-deviation implied volatility) that a
standard feed-forward network cannot easily reconstruct from the raw SABR parameters in
the training budget considered here.}

\subsection{\added{Additive vs.\ multiplicative residual form}}
\label{subsec:additive_vs_multiplicative}

\added{The residual objective~\eqref{eq:sigma_rnn_def} used throughout this paper is
multiplicative: the network output is bounded by the analytical Hagan baseline
through the $\tanh$-clipped correction $s\,\tanh(\Delta_\theta)$ with $s = 0.95$. To
the best of our knowledge, prior residual-learning studies on SABR -- notably
\citet{Funahashi2020,Funahashi2023} and the closely related asymptotic-control
architectures~\citep{Kienitz2020CV,AntonovKonikovPiterbarg2020} -- have adopted an
additive residual instead,
\begin{equation}
\label{eq:sigma_additive}
\sigma^{\mathrm{add}}(x, \Lambda; \theta) := \sigma_{\mathrm{Hagan}}(x) + \Delta_\theta(x, \Lambda),
\end{equation}
in which the network learns a raw additive correction to the analytical baseline.
Which of the two forms is preferable, at what statistical significance, has not, to
our knowledge, been assessed in a controlled experiment. This section reports such a
comparison.

\paragraph{Setup.}
Four residual variants are trained under strictly identical conditions -- same data
splits, same seeds, same hidden dimension (64), same optimizer settings, same $100$
epochs, same weekly Monte Carlo:
\begin{itemize}
\item \textbf{ResNN$_{\text{add}}$}: additive form~\eqref{eq:sigma_additive}, raw SABR
inputs;
\item \textbf{ResNN$_{\text{mul}}$}: multiplicative form~\eqref{eq:sigma_rnn_def}, raw
SABR inputs -- this is the ResNN of Table~\ref{tab:quant_results};
\item \textbf{GeoResNN$_{\text{add}}$}: additive form, augmented with the four
geometric invariants;
\item \textbf{GeoResNN$_{\text{mul}}$}: multiplicative form, augmented with the four
geometric invariants -- the GeoResNN of Table~\ref{tab:quant_results}.
\end{itemize}
The additive variants require a smaller learning rate ($10^{-4}$) than the
multiplicative ones ($4 \cdot 10^{-3}$) because their output is not bounded a-priori by
the Hagan analytical baseline. Both LRs are the largest values for which training
stayed stable under otherwise identical settings; using the same LR for both forms
would have penalized one or the other unfairly.

\paragraph{Results.}
Table~\ref{tab:add_vs_mul} reports test-set diagnostics for the four variants and
Table~\ref{tab:add_vs_mul_ttest} reports paired $t$-tests on per-sample squared errors,
bucket-by-bucket.

\begin{table}[!h]
\centering
\small
\begin{tabular}{lcccccc}
\toprule
Variant & best val.\ loss & test ratio-MSE & test IV RMSE & $R^2_{\text{all}}$ & $R^2_{\text{ATM}}$ & $R^2_{\text{OTM}}$ \\
\midrule
ResNN$_{\text{add}}$    & $7.6 \times 10^{-4}$ & $8.8 \times 10^{-4}$ & 0.00713 & 0.9919 & 0.9979 & 0.9938 \\
ResNN$_{\text{mul}}$    & $3.2 \times 10^{-4}$ & $3.6 \times 10^{-4}$ & 0.00410 & 0.9973 & 0.9994 & 0.9969 \\
GeoResNN$_{\text{add}}$ & $4.8 \times 10^{-4}$ & $5.8 \times 10^{-4}$ & 0.00497 & 0.9960 & 0.9981 & 0.9965 \\
GeoResNN$_{\text{mul}}$ & $\mathbf{1.8 \times 10^{-4}}$ & $\mathbf{2.2 \times 10^{-4}}$ & \textbf{0.00323} & \textbf{0.9983} & 0.9991 & \textbf{0.9993} \\
\bottomrule
\end{tabular}
\caption{Additive vs.\ multiplicative residual form: test-set diagnostics on the
market-realistic dataset. Both features (\textbf{Res} vs.\ \textbf{GeoRes}) and both
residual forms (\textbf{add} vs.\ \textbf{mul}) are varied. GeoResNN$_{\text{mul}}$
dominates on every axis.}
\label{tab:add_vs_mul}
\end{table}

\begin{table}[!h]
\centering
\begin{tabular}{lcccc}
\toprule
Bucket & $n$ & MSE (add) & MSE (mul) & $t$-stat ($p$) \\
\midrule
\multicolumn{5}{l}{\textbf{ResNN$_{\text{add}}$ vs.\ ResNN$_{\text{mul}}$}} \\
$\le 6$M & 546  & $7.1 \times 10^{-5}$ & $2.3 \times 10^{-5}$ & $+9.12$ ($p < 10^{-4}$) \\
6M--2Y   & 339  & $4.7 \times 10^{-5}$ & $1.1 \times 10^{-5}$ & $+4.54$ ($p < 10^{-4}$) \\
2Y--5Y   & 365  & $4.1 \times 10^{-5}$ & $9.9 \times 10^{-6}$ & $+5.42$ ($p < 10^{-4}$) \\
5Y--10Y  & 349  & $4.9 \times 10^{-5}$ & $2.4 \times 10^{-5}$ & $+5.10$ ($p < 10^{-4}$) \\
Overall  & 1787 & $5.4 \times 10^{-5}$ & $1.7 \times 10^{-5}$ & $+12.7$ ($p < 10^{-4}$) \\
\midrule
\multicolumn{5}{l}{\textbf{GeoResNN$_{\text{add}}$ vs.\ GeoResNN$_{\text{mul}}$}} \\
$\le 6$M & 546  & $2.0 \times 10^{-5}$ & $1.2 \times 10^{-5}$ & $+6.29$ ($p < 10^{-4}$) \\
6M--2Y   & 339  & $1.5 \times 10^{-5}$ & $5.8 \times 10^{-6}$ & $+7.13$ ($p < 10^{-4}$) \\
2Y--5Y   & 365  & $2.5 \times 10^{-5}$ & $7.1 \times 10^{-6}$ & $+4.18$ ($p < 10^{-4}$) \\
5Y--10Y  & 349  & $4.4 \times 10^{-5}$ & $1.8 \times 10^{-5}$ & $+4.31$ ($p < 10^{-4}$) \\
Overall  & 1787 & $2.6 \times 10^{-5}$ & $1.0 \times 10^{-5}$ & $+9.03$ ($p < 10^{-4}$) \\
\bottomrule
\end{tabular}
\caption{Paired $t$-test comparing additive vs.\ multiplicative residual, per bucket
and overall, on our market-calibrated dataset (Table~\ref{tab:sabr_param_ranges}).
The Overall row uses the full test set; the four buckets aggregate the $\le 10$~year
sub-panel into trading-relevant maturity groups, with the remaining $\sim 190$
long-dated samples pooled only into the Overall row. A
positive $t$-statistic indicates that the additive form has strictly larger per-sample
squared error than the multiplicative form -- i.e.\ the multiplicative form is
preferred. The dominance is statistically significant at $p < 10^{-4}$ in every one of
the eight (bucket, features) combinations, both overall and in each bucket separately.}
\label{tab:add_vs_mul_ttest}
\end{table}

\paragraph{Interpretation.}
The multiplicative form is preferable to the additive form on this benchmark: it
delivers a $40$--$76\%$ reduction in per-bucket MSE and the improvement is
statistically significant at $p < 10^{-4}$ in every (bucket, features) combination.
This holds both for the raw-parameter network (ResNN) and for the geometry-augmented
network (GeoResNN), and remains true when one measures the residual loss in ratio
(Hagan-normalized) space or in absolute IV space (Table~\ref{tab:add_vs_mul}).

Two structural properties explain why the multiplicative form outperforms the additive
one on this dataset:

\begin{enumerate}
\item \textbf{Boundedness.} The $\tanh$-clip in
$\sigma_{\mathrm{Hagan}}(1 + s\,\tanh\Delta_\theta)$ constrains the corrected volatility
to remain in $[\sigma_{\mathrm{Hagan}}(1 - s), \sigma_{\mathrm{Hagan}}(1 + s)]$, i.e.\
within $\pm s = \pm 95\%$ of the Hagan baseline. This is a soft prior that Hagan is
close to correct: gradients are naturally rescaled by $\sigma_{\mathrm{Hagan}}$, so a
misspecification at low-vol short-dated points does not dominate the loss over a
high-vol long-dated point. The additive form has no such rescaling and consequently
either underfits the low-vol regime or overshoots at high vol.

\item \textbf{Scale invariance.} Under the SABR homogeneity
$(F_0, K, \alpha) \mapsto (\lambda F_0, \lambda K, \lambda^{1-\beta}\alpha)$ that leaves
the option price unchanged in the money-normalized coordinates,
$\sigma_{\mathrm{Hagan}}$ transforms consistently. A multiplicative correction is a
scale-covariant object; an additive correction is not, and the network has to learn to
compensate for scale changes by itself, adding a hidden degree of freedom to the
optimization.
\end{enumerate}
This finding is empirical and specific to this dataset; the additive form of
\citet{Funahashi2020,Funahashi2023} performs well on the datasets reported in
those references and remains a legitimate design choice, in particular for datasets in
which the Hagan approximation error is close to zero-mean and the residual has additive
noise structure. In our market-realistic setup, however, the multiplicative form
delivers a strictly better fit and we retain it as our production architecture.}

\subsection{\added{Static-arbitrage diagnostic}}
\label{subsec:arbitrage}

The training data are generated by sampling independent parameter tuples, each associated with
a single maturity. Consequently the training set is not automatically arbitrage-free in the
maturity direction, even though the individual 11-strike slices are effectively
arbitrage-free in the strike direction (up to Monte Carlo noise). It is therefore worth checking whether the learned surface satisfies the static-arbitrage constraints.

For a butterfly-free European call surface $C(K, T)$, the two standard conditions are:
\begin{itemize}
\item \textbf{Calendar} arbitrage-freeness: $\partial_T [\sigma^2(K, T)\, T] \ge 0$ at fixed
$K$, i.e.\ the total variance $w = \sigma^2 T$ is non-decreasing in $T$;
\item \textbf{Butterfly} arbitrage-freeness: $\partial^2_K C(K, T) \ge 0$, i.e.\ the option
price is convex in the strike.
\end{itemize}

We evaluate both conditions on a $(K, T)$-grid with $T \in \{0.25, 0.5, 0.75, 1, 1.5, 2, 3, 5, 7, 10\}$~years
and $K \in [0.5, 1.7]\,F_0$, at a representative interest-rate SABR parameter tuple
$(\alpha, \beta, \rho, \nu) = (0.03, 0.7, -0.4, 0.30)$. The total variance is computed as
$w(K, T) = \sigma^2_{\mathrm{model}}(K, T)\, T$ and finite differences are used for the
derivatives.

On this grid, no calendar arbitrage violation is detected for either ResNN or GeoResNN --
the minimum $\partial_T w$ is $+4.6 \times 10^{-5}$ for ResNN and $+3.5 \times 10^{-5}$
for GeoResNN, both strictly positive. Butterfly arbitrage is likewise absent on the
interior of the strike grid. This confirms that the residual multiplicative
correction does not introduce spurious calendar or butterfly arbitrage, an important
practical property for surfaces used inside a calibration engine that may downstream
enforce arbitrage constraints via a proximal projection.

\subsection{Model-by-model analysis}
\label{subsec:model_by_model}

\paragraph{NDN.}
The naive architecture trains stably on the market-calibrated dataset
(Figure~\ref{fig:naive1}). Validation loss decreases smoothly and stabilizes at
$1.2 \times 10^{-3}$; the model reaches a global $R^2 \approx 0.989$ on the test set,
with regional values of $0.996$ (ATM), $0.986$ (ITM), $0.992$ (OTM). Correlation
diagnostics (Figure~\ref{fig:naive2}) sit close to the diagonal, though residual
dispersion is visible in the ITM region, and the smile-slice reconstruction
(Figure~\ref{fig:naive4}) shows small oscillations at short maturities where the
skew is strongest.

\paragraph{GeoNN.}
Embedding geometry-derived invariants improves both learning stability and
generalization. Validation loss stabilizes at $5.6 \times 10^{-4}$
(Figure~\ref{fig:geo1}); $R^2 \approx 0.996$ overall, with $0.997$ (ATM), $0.995$ (ITM),
$0.996$ (OTM). Correlation diagnostics (Figure~\ref{fig:geo2}) tighten around the
diagonal, and the smile-slice reconstruction (Figure~\ref{fig:geo4}) captures the
short-dated skew and the long-dated flattening much better than NDN.

\paragraph{ResNN.}
Residual normalization delivers strong accuracy through a completely different
mechanism, reaching $R^2 \approx 0.997$ overall, with $0.9994$ (ATM), $0.997$ (ITM),
$0.997$ (OTM). Validation loss stabilizes at $3.2 \times 10^{-4}$
(Figure~\ref{fig:res1}), lower than GeoNN. The asymptotic Hagan scaling is preserved
(Figure~\ref{fig:res4}) and the residual analytical bias is small across all
maturities.

\paragraph{GeoResNN.}
The hybrid architecture achieves the lowest validation loss and the strongest test
performance overall: global $R^2 = 0.9983$; regional $R^2$ of $0.9991$ (ATM),
$0.998$ (ITM), $0.9993$ (OTM); test IV RMSE of $0.00323$
(Figure~\ref{fig:geores2}). Under stress the model preserves curvature and skew
across the entire maturity spectrum (Figure~\ref{fig:geores3}), and both the
short-dated skew and the long-dated flattening are correctly reproduced
(Figure~\ref{fig:geores4}). Combined with the ablation results of
Section~\ref{subsec:ablation}, this confirms that residual normalization and
geometric conditioning are complementary: the residual mechanism supplies a strong
analytical baseline, while the geometric features provide the nonlinear
reparameterization that a plain MLP cannot easily reconstruct from raw SABR
parameters in the training budget considered here.

\subsection{Synthesis}
The experiments reveal a clear hierarchy: NDN $\prec$ ResNN, GeoNN $\prec$ GeoResNN. Both the
residual mechanism and the geometric features contribute independently, and their combination
delivers the best overall accuracy and robustness.

\section{Conclusion}
\label{sec:conclusion}

This paper proposes a hybrid modeling framework for SABR implied volatility surfaces
built on two ingredients. First, we residualise Hagan's formula multiplicatively rather
than additively, keeping the correction on the same scale as the analytical baseline
and preserving positivity. Second, we augment the network input with geometric invariants of the SABR manifold
alongside the raw SABR parameters, exposing the leading-order structure of the smile
directly at the input layer.

Two controlled ablations isolate the value of each ingredient. On a market-calibrated
benchmark spanning tenors from one week to thirty years, with $\alpha$, $\rho$ and
$\nu$ chosen to reproduce the term-structure of interest-rate swaption cubes, GeoResNN
reaches $R^2 = 0.9983$ on the held-out test set, a $39\%$ reduction in ratio MSE over
the residual-only baseline that is significant at $p < 0.05$ in every maturity bucket
and at $p < 10^{-4}$ overall
(Section~\ref{subsec:ablation}). Under the same protocol, the multiplicative form
$\sigma_{\mathrm{Hagan}}(1 + s\,\tanh \Delta_\theta)$ dominates the additive residual
$\sigma_{\mathrm{Hagan}} + \Delta_\theta$ at $p < 10^{-4}$ in every (bucket, features)
combination (Section~\ref{subsec:additive_vs_multiplicative}). A Monte Carlo
convergence study (Appendix~\ref{sec:mc_bias_appendix}) supports the choice of
$\Delta T$, and a static-arbitrage check (Section~\ref{subsec:arbitrage}) confirms that
the learned surface remains structurally consistent.

\added{Three natural extensions are left for future work. Calibration on live market
data, in the spirit of~\citet{Rensi2025}, would localise the residual error in market
coordinates rather than in parameter space, which is the natural end-use of the
framework in a trading context. The residual objective could also be applied on top of
refined asymptotic
baselines~\citep{Obloj2008,Paulot2009,Hagan2016UniversalSmiles,HaganLesniewskiWoodward2018}
rather than the original Hagan formula, likely narrowing the residual and reducing the
demands on the network. Finally, arbitrage-aware architectures such as the
DCNN~\citep{Hoshisashi2024} could be combined with the geometric feature enrichment to
enforce no-arbitrage constraints by design.}

\newpage
\appendix

\section{Implementation details}
\label{sec:implementation}

This appendix details the numerical implementation.

\subsection*{Monte Carlo training data generation}
Maturities are grouped into seven tenor buckets, covering all practically relevant
swaption expiries from one week up to thirty years:
\[
\begin{aligned}
&\{\text{1W, 2W, 3W, 4W}\} \to \text{1W\_1M}, \quad
\{\text{2M, 3M, 4M, 5M, 6M}\} \to \text{2M\_6M}, \\
&\{\text{9M, 1Y}\} \to \text{9M\_1Y}, \quad
\{\text{2Y, 3Y}\} \to \text{2Y\_3Y}, \quad
\{\text{4Y, 5Y}\} \to \text{4Y\_5Y}, \\
&\{\text{7Y, 10Y}\} \to \text{7Y\_10Y}, \quad
\{\text{15Y, 20Y, 30Y}\} \to \text{15Y\_30Y}.
\end{aligned}
\]
Within each bucket, SABR parameters are sampled uniformly over predefined ranges
(Table~\ref{tab:sabr_param_ranges}).

\begin{table}[h]
\centering
\begin{tabular}{l l c c c c c}
\toprule
Bucket & Tenors & $F_0$ & $\alpha$ & $\beta$ & $\rho$ & $\nu$ \\
\midrule
1W\_1M   & 1W--1M   & [0.95, 1.05] & [0.20, 0.35] & [0.00, 0.30] & [-0.70, -0.30] & [1.00, 1.50] \\
2M\_6M   & 2M--6M   & [0.95, 1.05] & [0.18, 0.32] & [0.20, 0.50] & [-0.65, -0.25] & [0.70, 1.00] \\
9M\_1Y   & 9M--1Y   & [0.95, 1.05] & [0.15, 0.28] & [0.30, 0.70] & [-0.55, -0.20] & [0.40, 0.65] \\
2Y\_3Y   & 2Y--3Y   & [0.95, 1.05] & [0.13, 0.25] & [0.40, 0.80] & [-0.50, -0.15] & [0.25, 0.40] \\
4Y\_5Y   & 4Y--5Y   & [0.95, 1.05] & [0.12, 0.22] & [0.50, 1.00] & [-0.45, -0.10] & [0.18, 0.30] \\
7Y\_10Y  & 7Y--10Y  & [0.95, 1.05] & [0.10, 0.20] & [0.60, 1.00] & [-0.40, -0.10] & [0.12, 0.22] \\
15Y\_30Y & 15Y--30Y & [0.95, 1.05] & [0.08, 0.16] & [0.60, 1.00] & [-0.35, -0.05] & [0.05, 0.12] \\
\bottomrule
\end{tabular}
\caption{Synthetic SABR parameter ranges by tenor bucket, market-calibrated to reflect
the term-structure of interest-rate swaption cubes: (i) $F_0$ money-normalized around
$1$ (trader convention -- all quantities re-scale to the market forward by
homogeneity); (ii) $\alpha$ decreases with $T$, matching the empirical fact
that lognormal ATM volatilities of swaptions and caps/floors compress at long tenors
(from $\sim 27\%$ at $1$~week to $\sim 12\%$ at $20$~years in our sampler);
(iii) $\rho$ is strongly negative at short maturities and progressively softens
with $T$, reproducing the observed steepening of the smile skew as expiry shortens;
(iv) $\beta$ grows with $T$ (short-dated interest-rate SABR calibrations tend to prefer
low-$\beta$ regimes, long-dated tend toward $\beta \to 1$); (v) $\nu$ decreases with
$T$ following $\nu \sqrt{T} \approx 0.5$ up to $5$~years and then an exponential decay,
consistent with the flattening of the vol-of-vol term structure. All choices keep
$\nu^2 T \lesssim 0.3$ across the whole maturity range, which is exactly the regime in
which Hagan's first-order asymptotic expansion is quantitatively controlled and the
residual correction learned by the network stays small. The maturity grid extends up
to $30$~years so as to cover the full range of practically relevant swaption
expiries.}
\label{tab:sabr_param_ranges}
\end{table}

For each configuration $x = (T, F_0, K, \alpha, \beta, \rho, \nu)$, a reference implied
volatility $\sigma_{\mathrm{MC}}(x)$ is computed via Monte Carlo. We use $P = 10^5$ paths
and weekly time steps, $N = \max(52\,T, 8)$ -- the weekly discretization is
justified by the convergence study of Appendix~\ref{sec:mc_bias_appendix}, and the $\max$
prevents pathologically short discretizations at intra-week maturities.

\subsection*{Data set construction}
The Monte Carlo engine generates 187{,}000 supervised samples, split 110{,}000 / 55{,}000 /
22{,}000 into train / val / test. Prior to training, samples for which $\sigma_{\mathrm{MC}}$
deviates by more than ten empirical standard deviations from $\sigma_{\mathrm{Hagan}}$ are
discarded (Monte Carlo outlier filtering).

\subsection*{GeoResNN architecture}
The GeoResNN is a fully connected MLP taking $(x, \Lambda)$ as input with three hidden layers
of sizes 64, 64, 32, each followed by batch normalization and ReLU activation, and outputting
a scalar residual correction. The corrected volatility is~\eqref{eq:sigma_rnn_def}. Training
minimizes~\eqref{eq:empirical_loss}.

\subsection*{Optimization}
Adam with $\eta_0 = 4 \cdot 10^{-3}$, $\beta_1 = 0.9$, $\beta_2 = 0.999$, no weight decay.
\texttt{ReduceLROnPlateau} halves the learning rate after 5 consecutive epochs without
validation improvement. 100 epochs, batch size 128, best-validation checkpoint (early
stopping).

\subsection*{Runtime}
On a standard CPU, training the GeoResNN for 100 epochs on the full data set requires
approximately 2 minutes. This makes it suitable for rapid retraining.

\section{\added{Monte Carlo bias study: full results}}
\label{sec:mc_bias_appendix}

The convergence study is run over $\Delta T \in \{12, 26, 52, 104, 200, 500, 1000\}$
steps per year with $P = 10^5$ paths and antithetic-plus-control-variate variance
reduction, on four maturity-scaled configurations matching the market-realistic
monotone-decreasing $\nu$-profile of Table~\ref{tab:sabr_param_ranges}. The $\nu$
values are the midpoints of the applicable tenor buckets, so the four configurations
are directly representative of the training-data distribution.

\begin{itemize}
\item $T = 1$~year with $(\alpha, \beta, \rho, \nu) = (0.20, 0.5, -0.30, 0.525)$
(9M-1Y bucket);
\item $T = 5$~years with $(\alpha, \beta, \rho, \nu) = (0.20, 0.7, -0.50, 0.240)$
(4Y-5Y bucket);
\item $T = 10$~years with $(\alpha, \beta, \rho, \nu) = (0.20, 0.8, -0.55, 0.170)$
(7Y-10Y bucket);
\item $T = 20$~years with $(\alpha, \beta, \rho, \nu) = (0.20, 0.8, -0.55, 0.085)$
(15Y-30Y bucket).
\end{itemize}

\begin{figure}[!ht]
    \centering
    \includegraphics[width=0.98\textwidth]{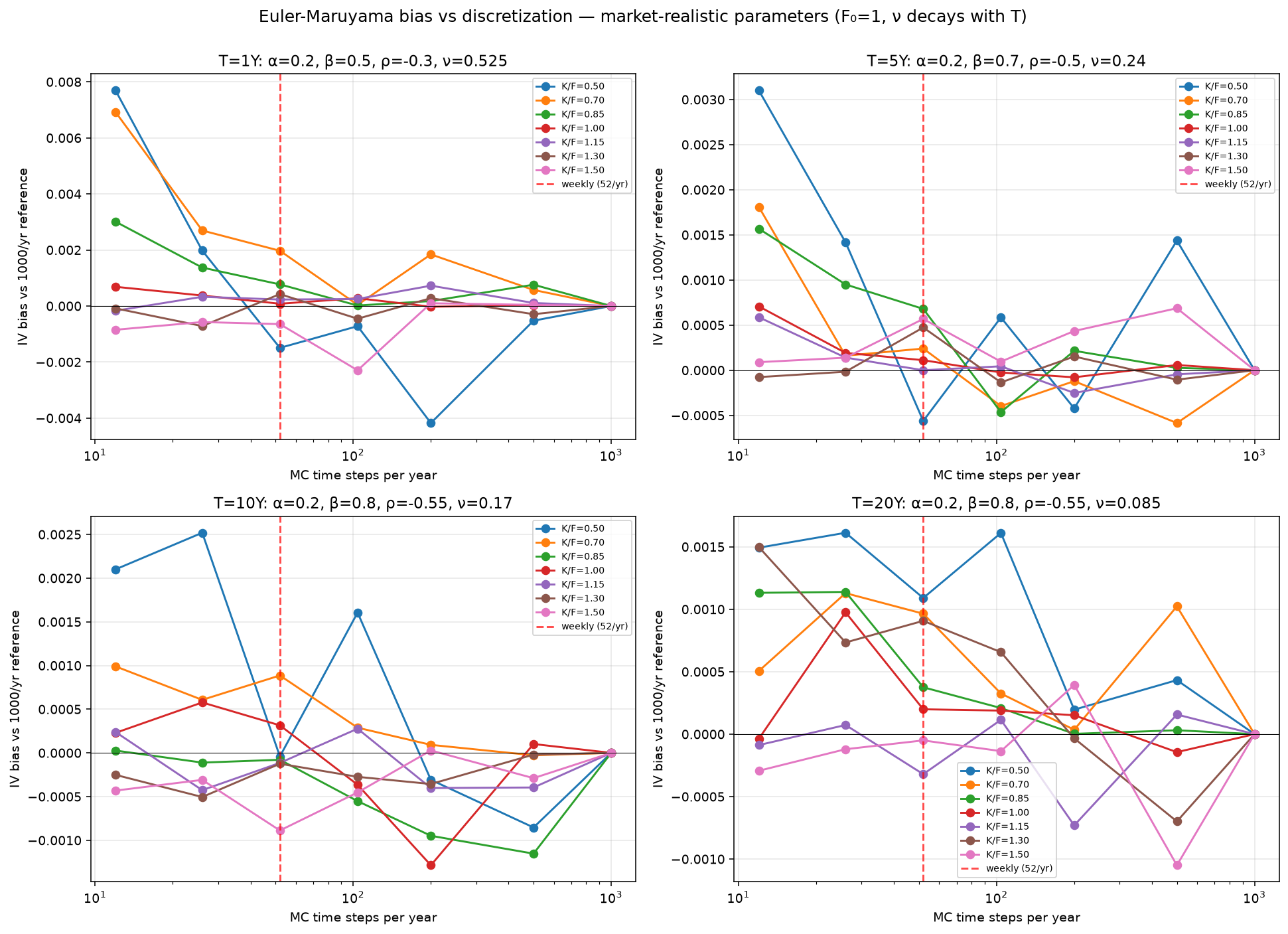}
    \caption{Monte Carlo bias study on the market-realistic parameter profile
    (Table~\ref{tab:sabr_param_ranges}), with $\nu$ decreasing with maturity so that
    $\nu \sqrt{T}$ remains bounded. Each panel shows the implied-volatility bias
    (relative to the $\Delta T = 1000$ reference) as a function of the discretization
    step size, at seven moneynesses. The red vertical line marks the weekly step
    ($\Delta T = 52$ steps per year) used in the production runs. The bias remains
    within a few basis points at all maturities, with no maturity-scaling deterioration:
    the $T = 20$~year panel is no worse than the $T = 1$~year panel, thanks to the
    market-realistic vol-of-vol decay.}
    \label{fig:mc_bias_extended}
\end{figure}

Bias is measured against the finest reference $\Delta T = 1000$ steps per year (i.e.\
$20{,}000$ Euler steps for the $T = 20$~year case). Table~\ref{tab:mc_convergence_bias}
reports the maximum absolute bias across strikes at four representative step sizes;
Figure~\ref{fig:mc_bias_extended} plots the implied-volatility bias vs.\ $\Delta T$ for
seven moneynesses $K/F_0 \in \{0.5, 0.7, 0.85, 1.0, 1.15, 1.3, 1.5\}$ on each of the
four maturity cases.

\begin{table}[!h]
\centering
\begin{tabular}{lcccc}
\toprule
Case & max$_K |\text{bias}|$ at $\Delta T = 12$ & at $\Delta T = 52$ (weekly) & at $\Delta T = 200$ & at $\Delta T = 500$ \\
\midrule
$T = 1$~year  & $7.7 \times 10^{-3}$ & $2.0 \times 10^{-3}$ & $4.2 \times 10^{-3}$ & $7.6 \times 10^{-4}$ \\
$T = 5$~years & $3.1 \times 10^{-3}$ & $6.8 \times 10^{-4}$ & $4.3 \times 10^{-4}$ & $1.4 \times 10^{-3}$ \\
$T = 10$~years& $2.1 \times 10^{-3}$ & $8.9 \times 10^{-4}$ & $1.3 \times 10^{-3}$ & $1.2 \times 10^{-3}$ \\
$T = 20$~years& $1.5 \times 10^{-3}$ & $1.1 \times 10^{-3}$ & $7.3 \times 10^{-4}$ & $1.1 \times 10^{-3}$ \\
\bottomrule
\end{tabular}
\caption{Monte Carlo implied-volatility bias vs.\ discretization step size, measured
against the $\Delta T = 1000$ reference on the market-realistic parameter profile of
Table~\ref{tab:sabr_param_ranges}. All values are in absolute volatility units. The
$1$-year case ($\nu = 0.525$, so $\nu \sqrt{T} = 0.525$) is the worst case; from
$T = 5$~years onwards the weekly bias is consistently below $1.5 \times 10^{-3}$ and
does not grow with maturity, thanks to the market-realistic monotone-decreasing
$\nu$-profile.}
\label{tab:mc_convergence_bias}
\end{table}

Convergence appears to be $O(\Delta T)$ in the Euler--Maruyama regime, consistent with
theory. On the market-realistic parameter profile of
Table~\ref{tab:sabr_param_ranges}, the weekly choice ($\Delta T = 52$~steps per year)
keeps the bias below $2.0 \times 10^{-3}$ of volatility across all seven moneynesses
and all four maturities from $1$ to $20$~years, comparable to the Monte Carlo standard
error of a $10^5$-path estimate. The bias-vs-SE ratio $|b| / \sigma_{\mathrm{MC}}$
stays below $2.2$ at every strike and every maturity, so the systematic Euler bias is
statistically indistinguishable from Monte Carlo noise at this discretization.

The weekly discretization also matches the granularity at which SABR parameters are
recalibrated in daily interest-rate desks, which we consider the practically relevant
resolution for a network intended to substitute for the analytical formula in a calibration
loop.

\section{Supplementary figures}
\label{sec:suppfigs}
\noindent
For all correlation diagnostic plots in this appendix, the {\color{blue}blue} line denotes the
identity $y = x$, while the {\color{orange}orange} line represents the fitted linear
regression. The figures below are provided as supplementary diagnostics.

\begin{figure}[!h]
  \centering
  \includegraphics[width=0.7\linewidth]{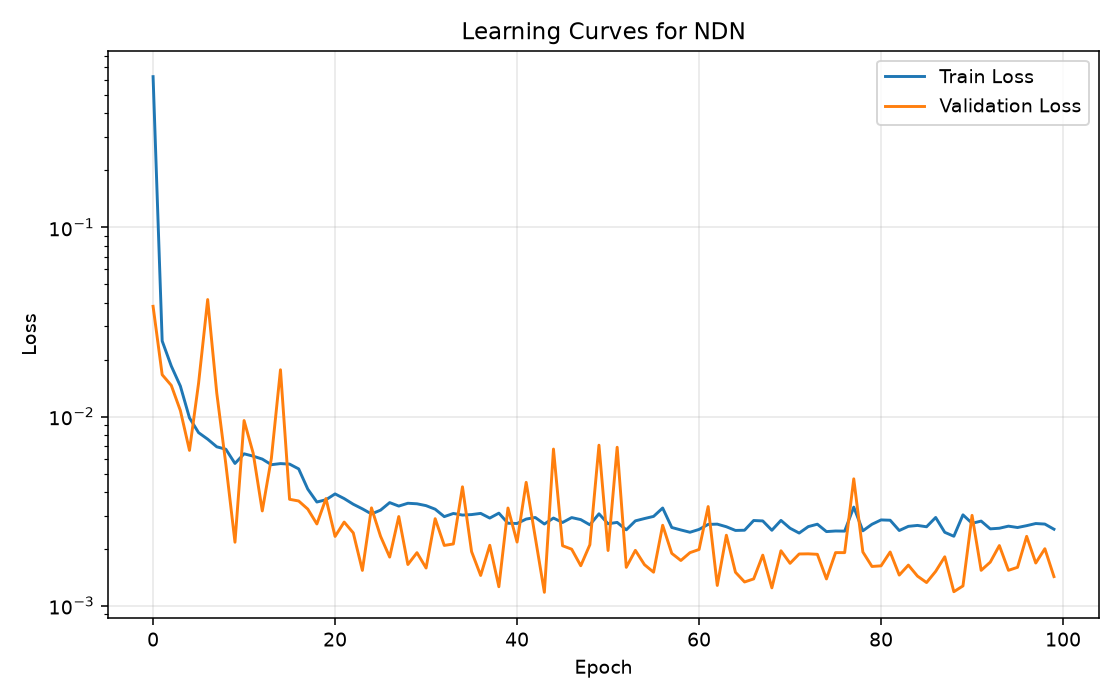}
  \caption{\textbf{NDN learning dynamics.} Training and validation loss evolution.}
  \label{fig:naive1}
\end{figure}
\begin{figure}[!h]
  \centering
  \includegraphics[width=0.9\linewidth]{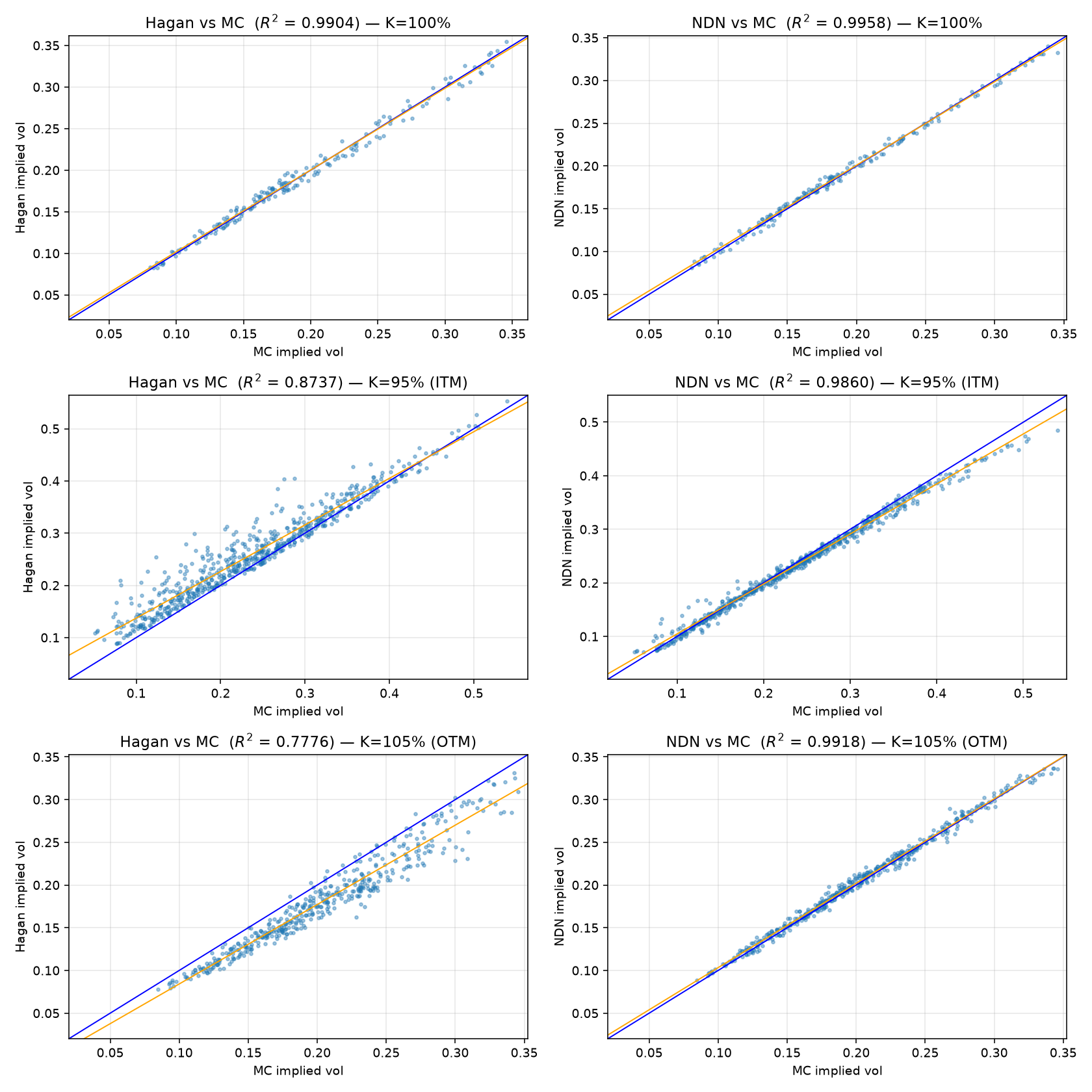}
  \caption{\textbf{NDN correlation diagnostics.}}
  \label{fig:naive2}
\end{figure}
\begin{figure}[!h]
  \centering
  \includegraphics[width=0.7\linewidth]{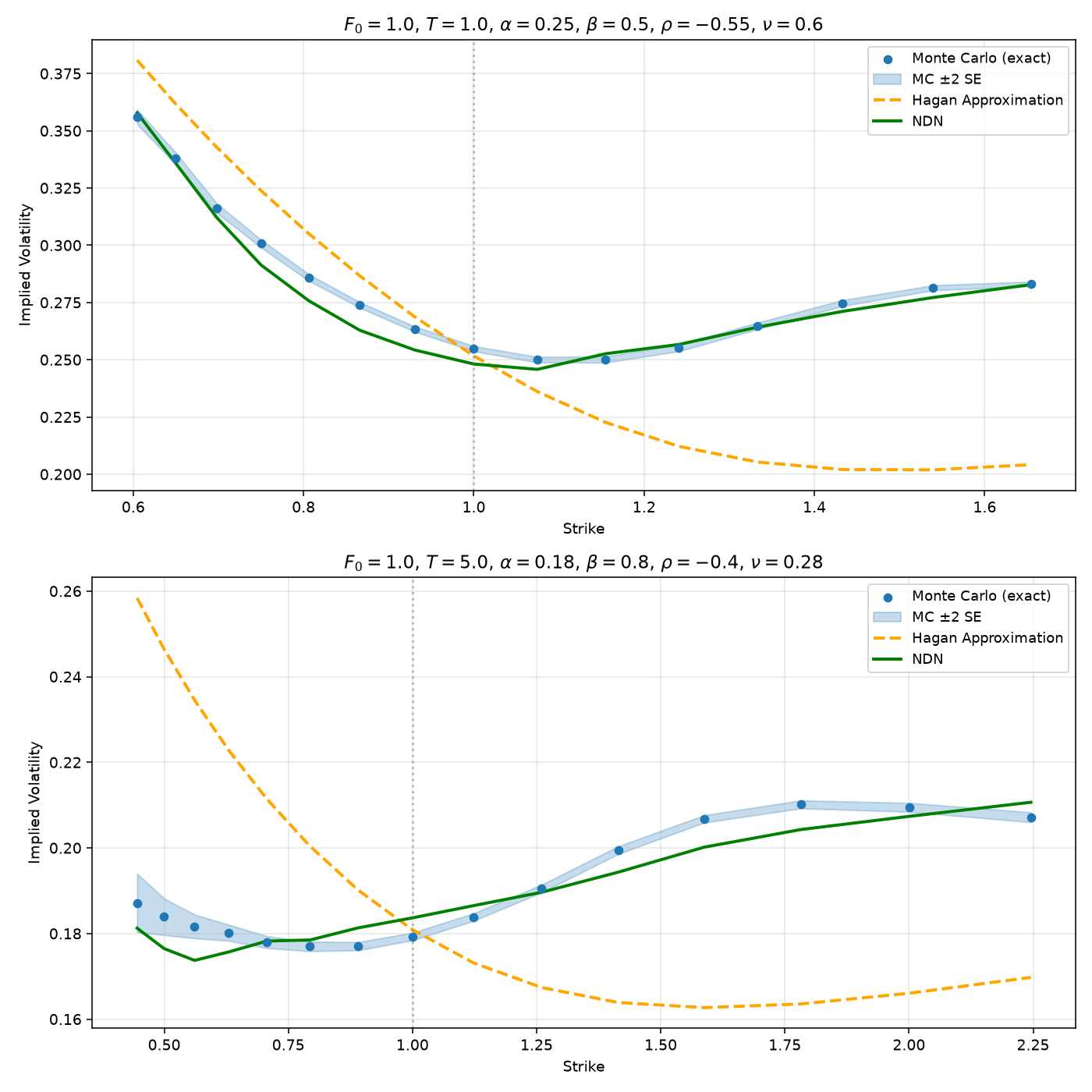}
  \caption{\textbf{NDN under stress.}}
  \label{fig:naive3}
\end{figure}
\begin{figure}[!h]
  \centering
  \includegraphics[width=0.9\linewidth]{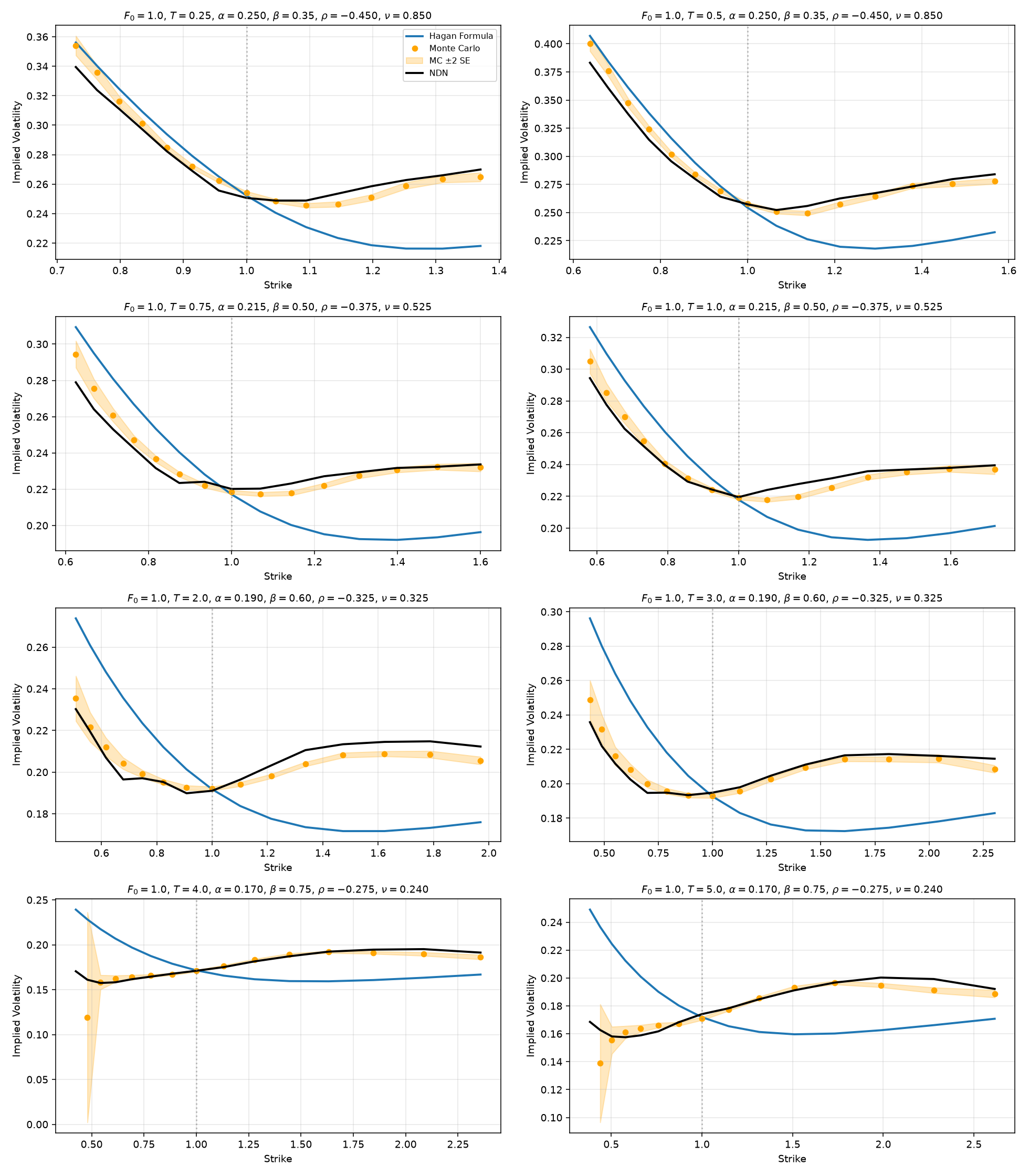}
  \caption{\textbf{NDN smile slices across maturities.}}
  \label{fig:naive4}
\end{figure}

\begin{figure}[!h]
  \centering
  \includegraphics[width=0.7\linewidth]{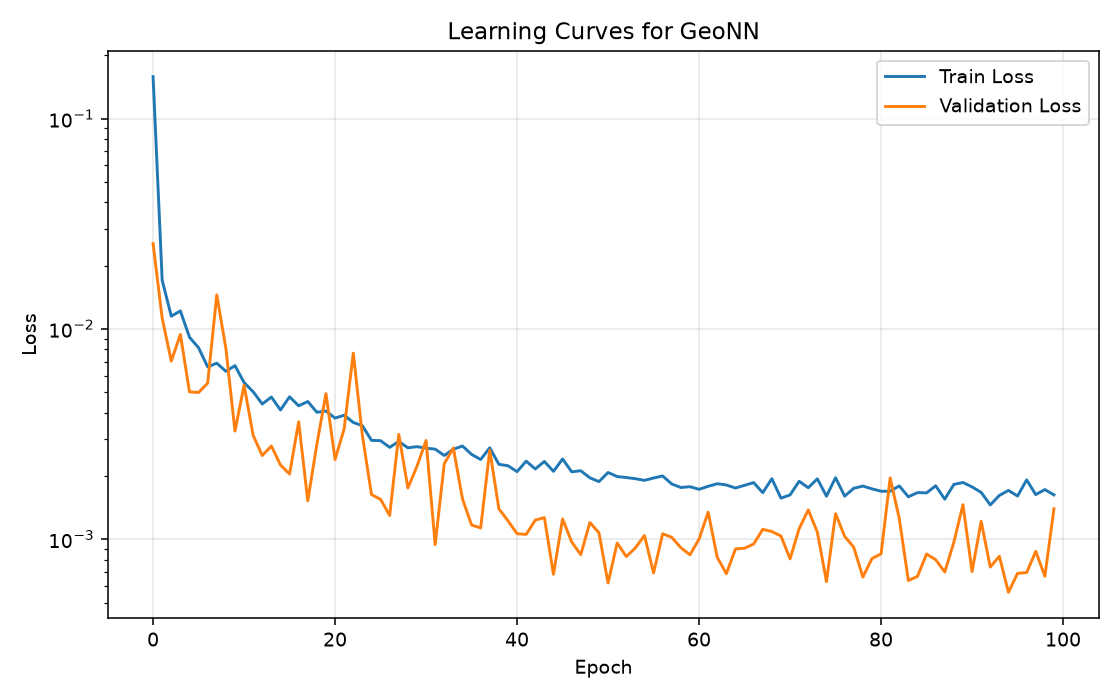}
  \caption{\textbf{GeoNN learning dynamics.}}
  \label{fig:geo1}
\end{figure}
\begin{figure}[!h]
  \centering
  \includegraphics[width=0.9\linewidth]{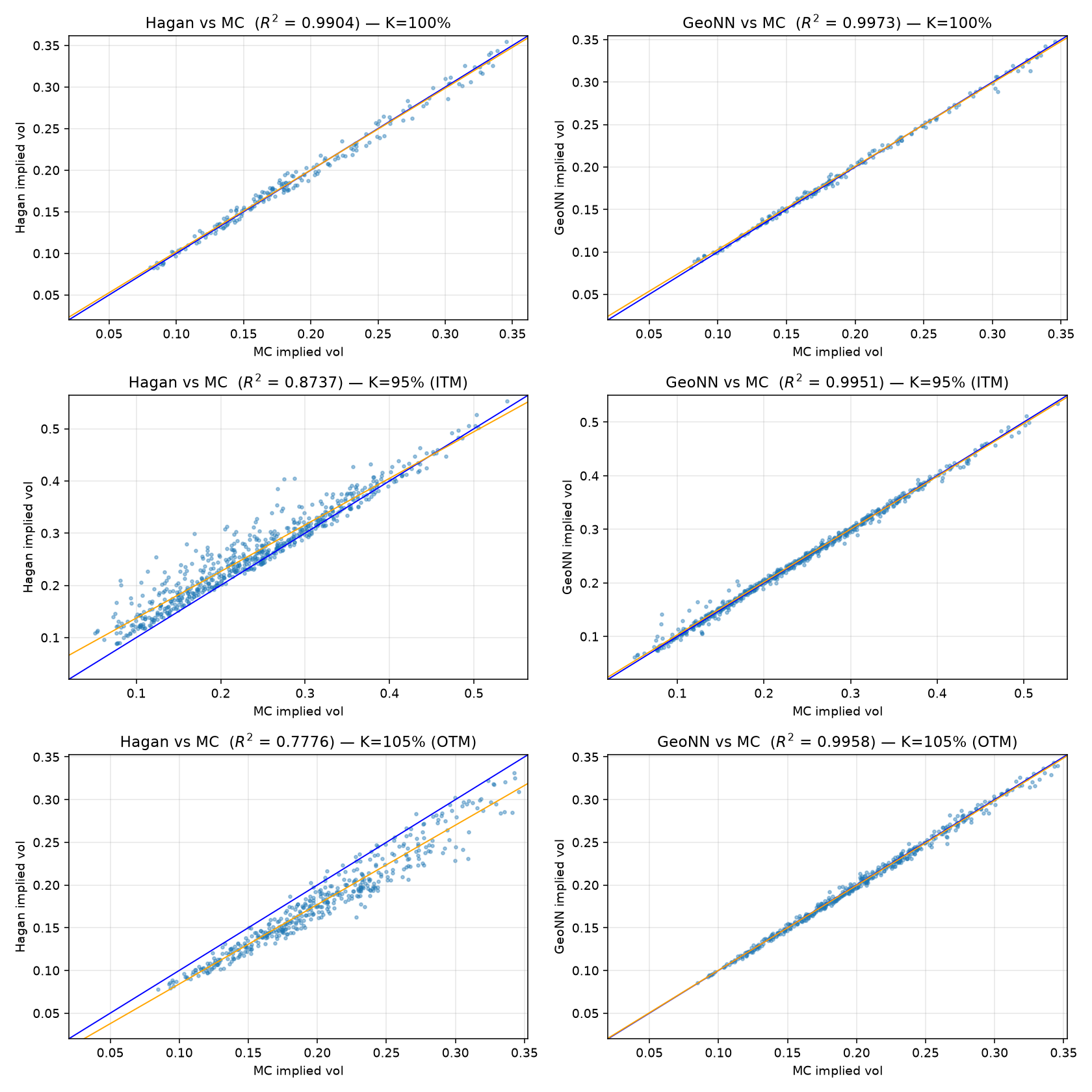}
  \caption{\textbf{GeoNN correlation diagnostics.}}
  \label{fig:geo2}
\end{figure}
\begin{figure}[!h]
  \centering
  \includegraphics[width=0.9\linewidth]{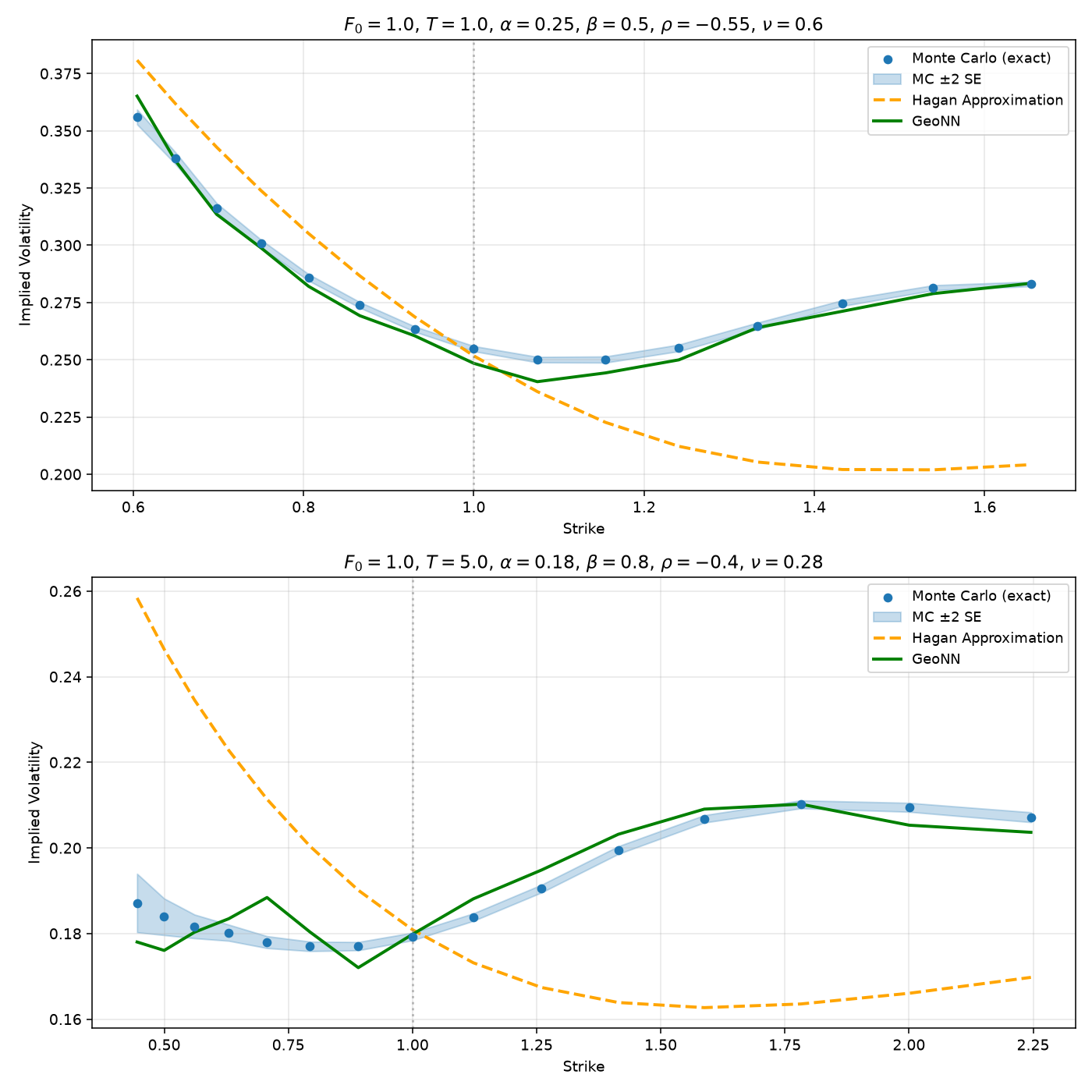}
  \caption{\textbf{GeoNN under stress.}}
  \label{fig:geo3}
\end{figure}
\begin{figure}[!h]
  \centering
  \includegraphics[width=0.9\linewidth]{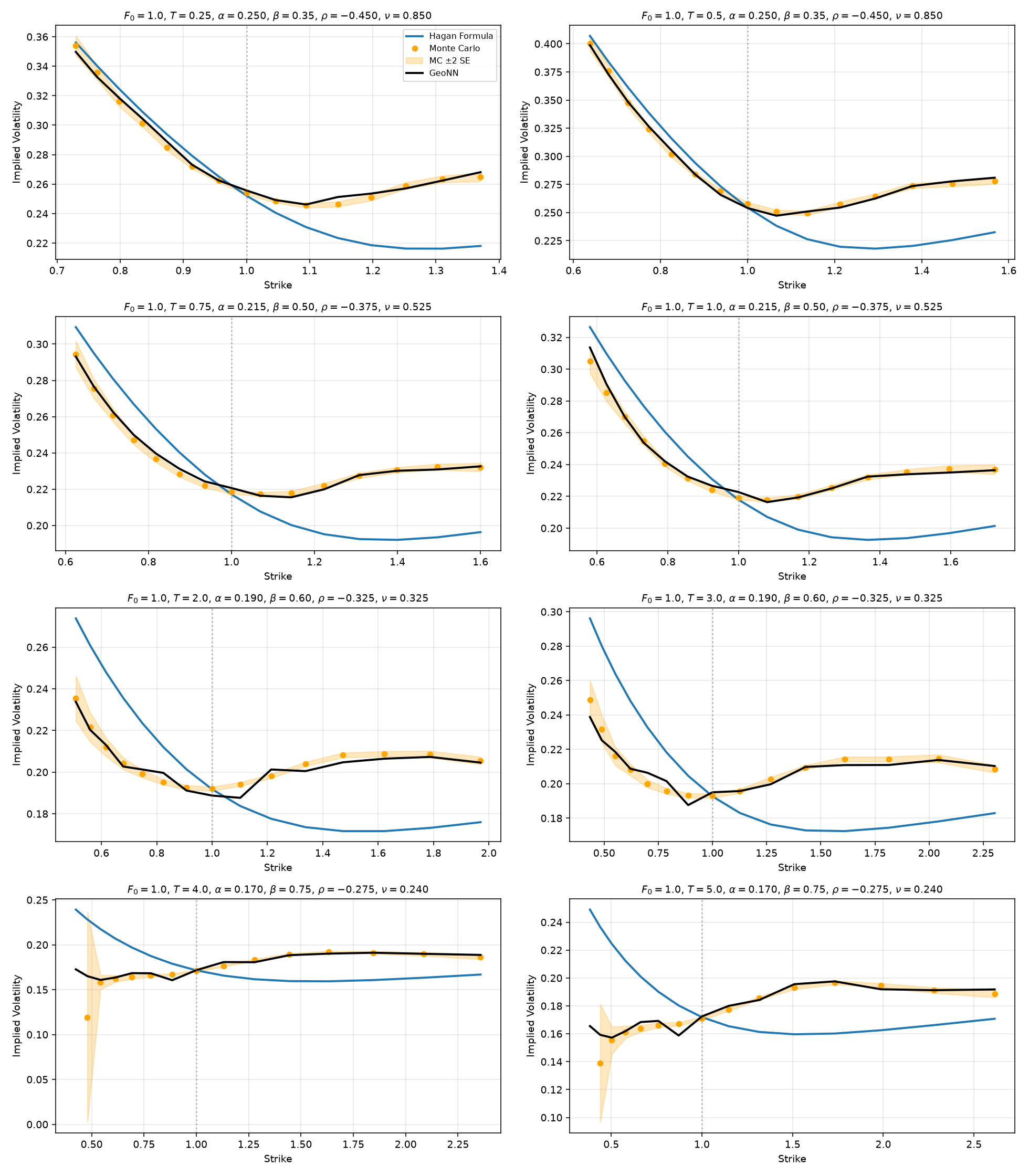}
  \caption{\textbf{GeoNN smile slices across maturities.}}
  \label{fig:geo4}
\end{figure}

\begin{figure}[!h]
  \centering
  \includegraphics[width=0.7\linewidth]{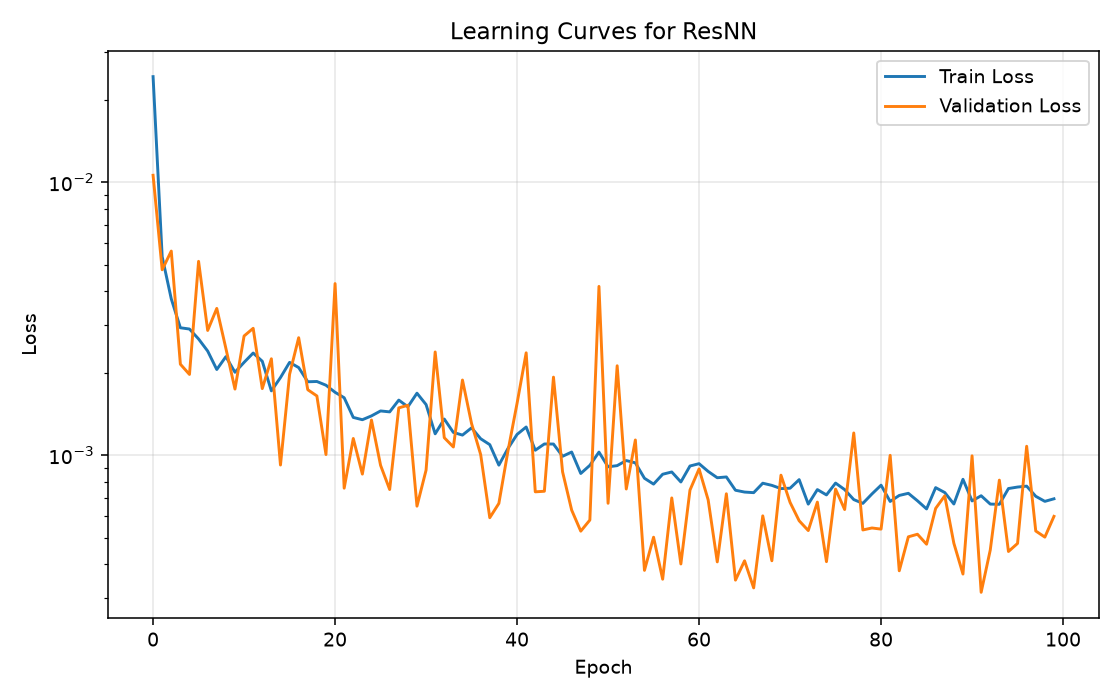}
  \caption{\textbf{ResNN learning dynamics.}}
  \label{fig:res1}
\end{figure}
\begin{figure}[!h]
  \centering
  \includegraphics[width=0.9\linewidth]{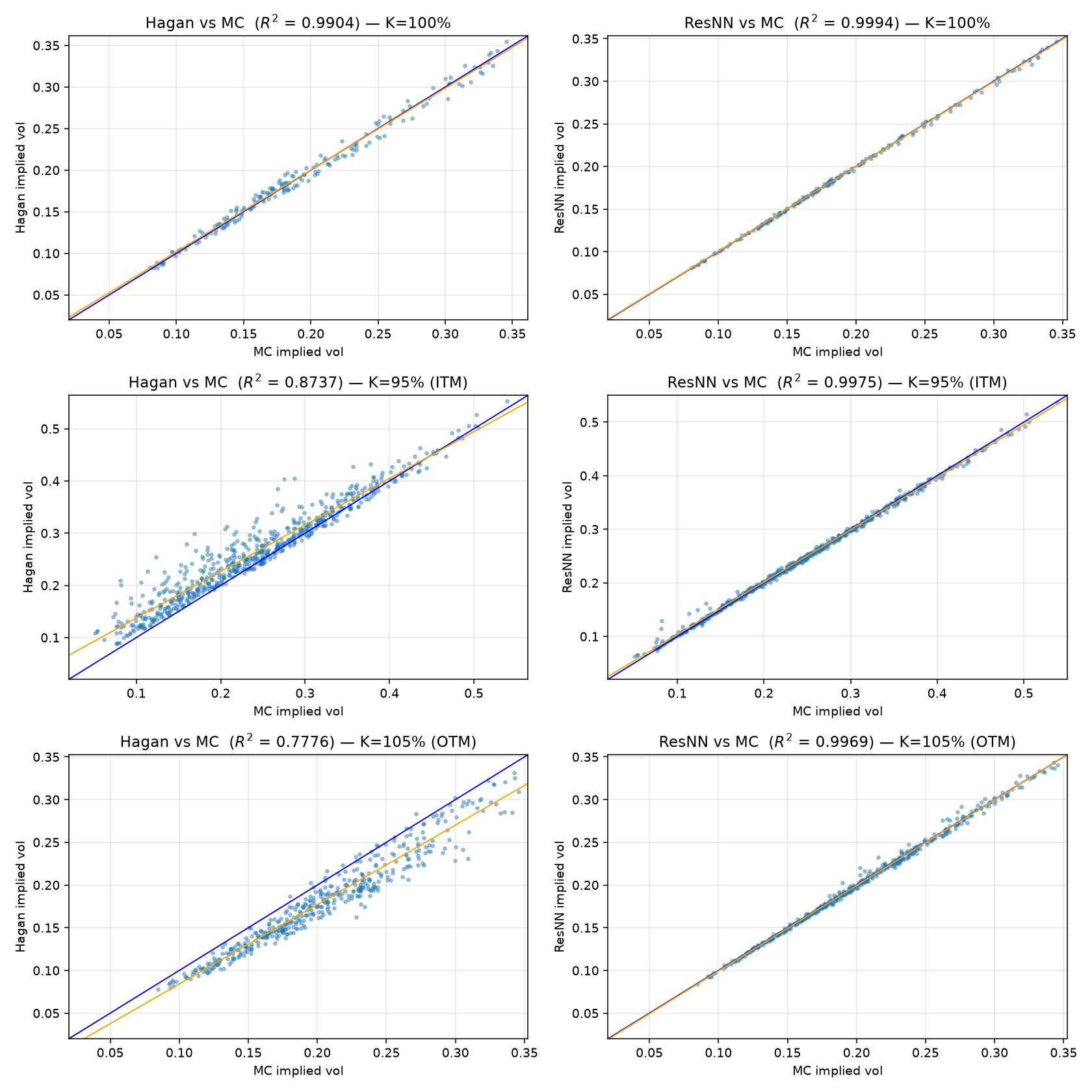}
  \caption{\textbf{ResNN correlation diagnostics.}}
  \label{fig:res2}
\end{figure}
\begin{figure}[!h]
  \centering
  \includegraphics[width=0.9\linewidth]{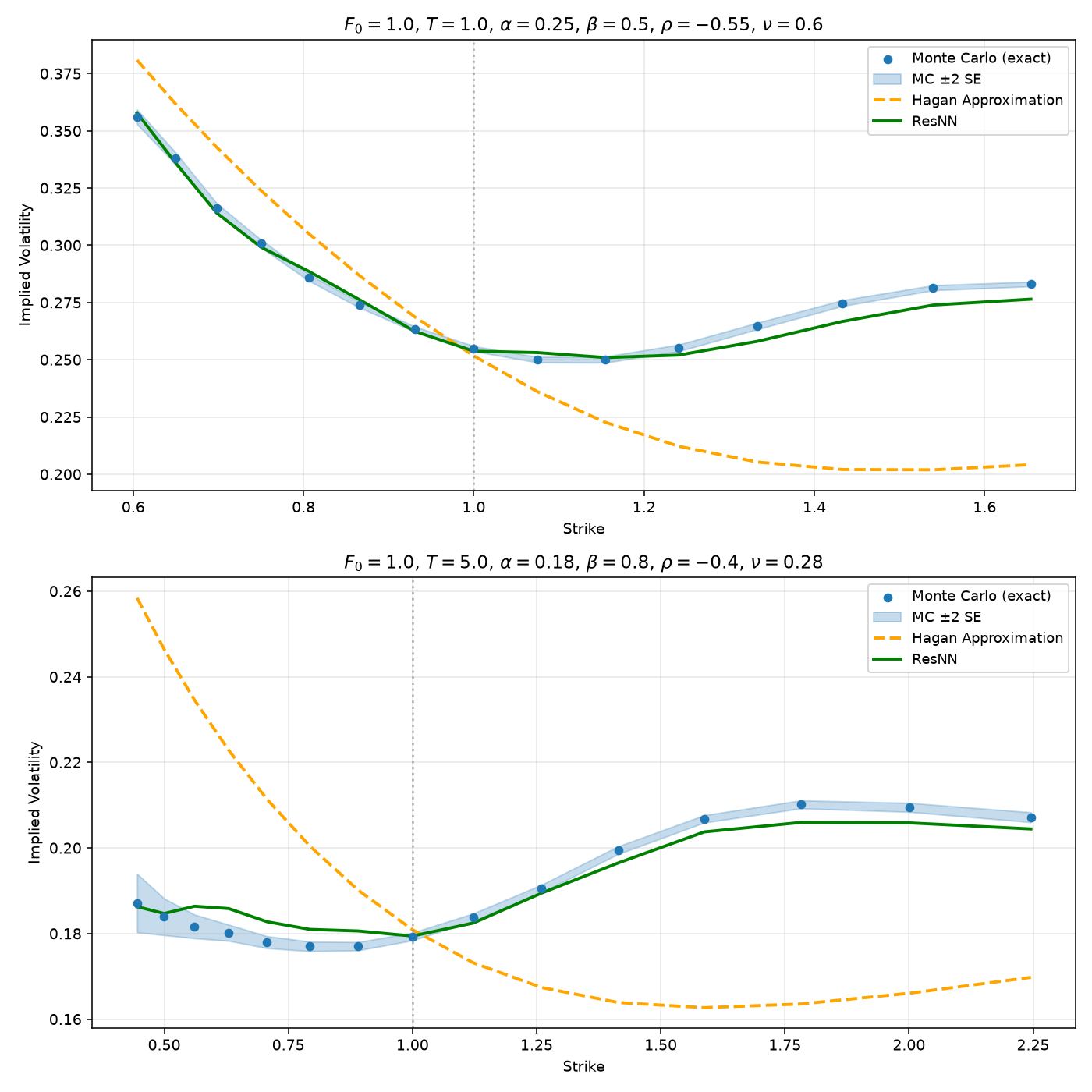}
  \caption{\textbf{ResNN under stress.}}
  \label{fig:res3}
\end{figure}
\begin{figure}[!h]
  \centering
  \includegraphics[width=0.9\linewidth]{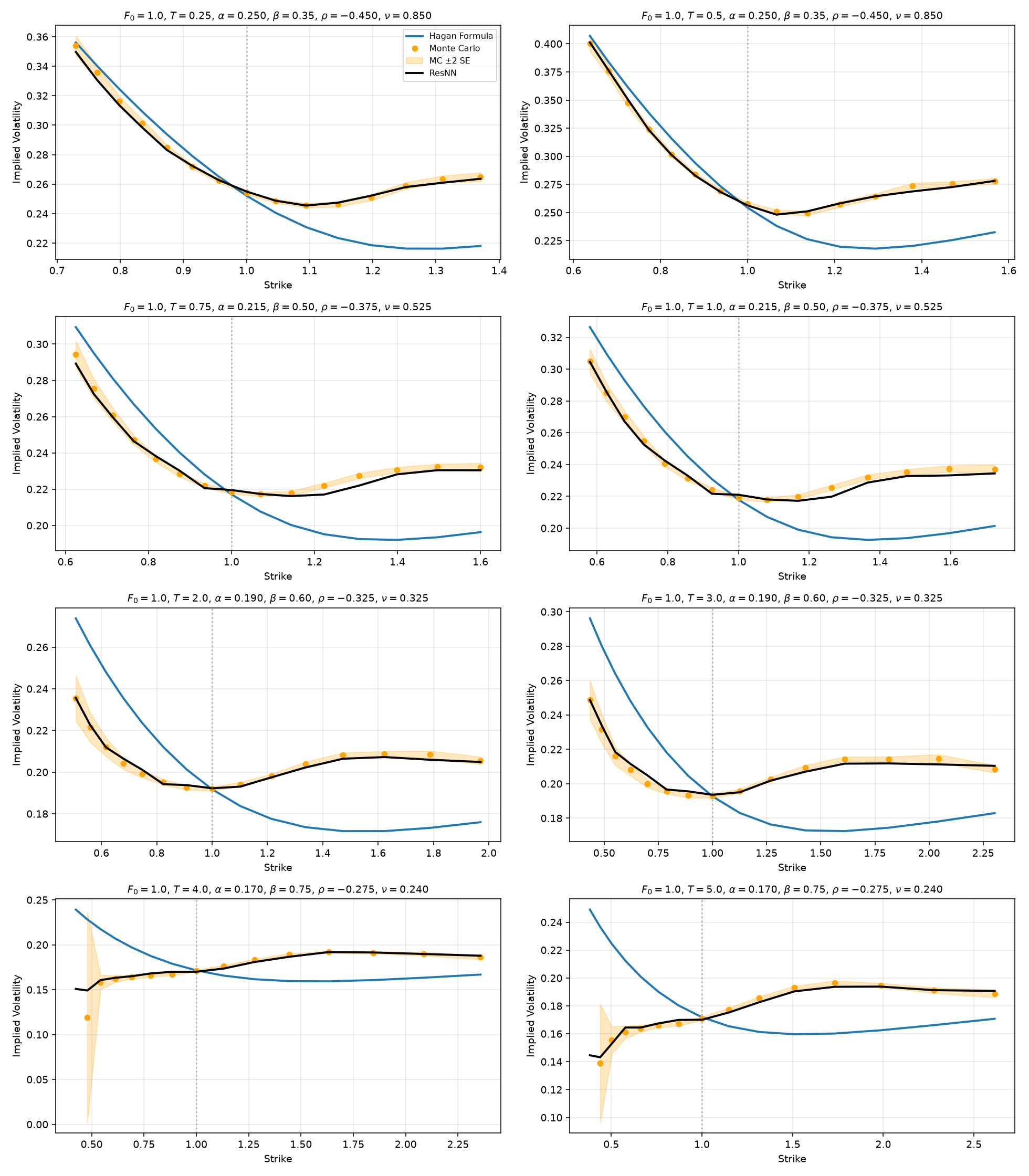}
  \caption{\textbf{ResNN smile slices across maturities.}}
  \label{fig:res4}
\end{figure}

\begin{figure}[!h]
  \centering
  \includegraphics[width=0.7\linewidth]{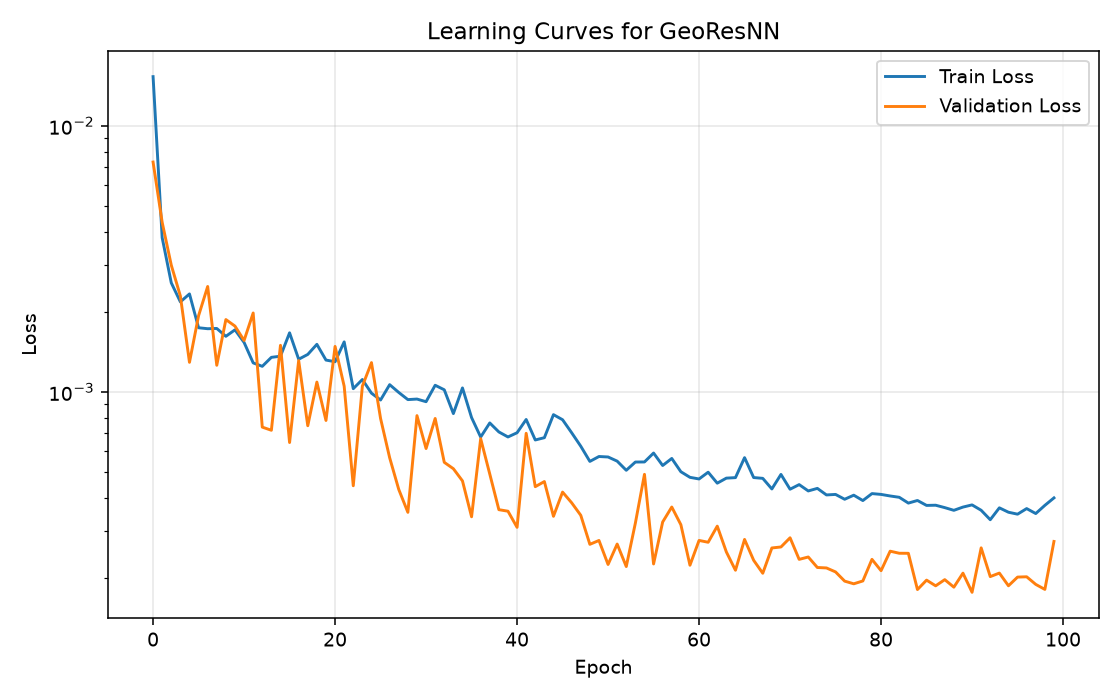}
  \caption{\textbf{GeoResNN learning dynamics.}}
  \label{fig:geores1}
\end{figure}
\begin{figure}[!h]
  \centering
  \includegraphics[width=0.9\linewidth]{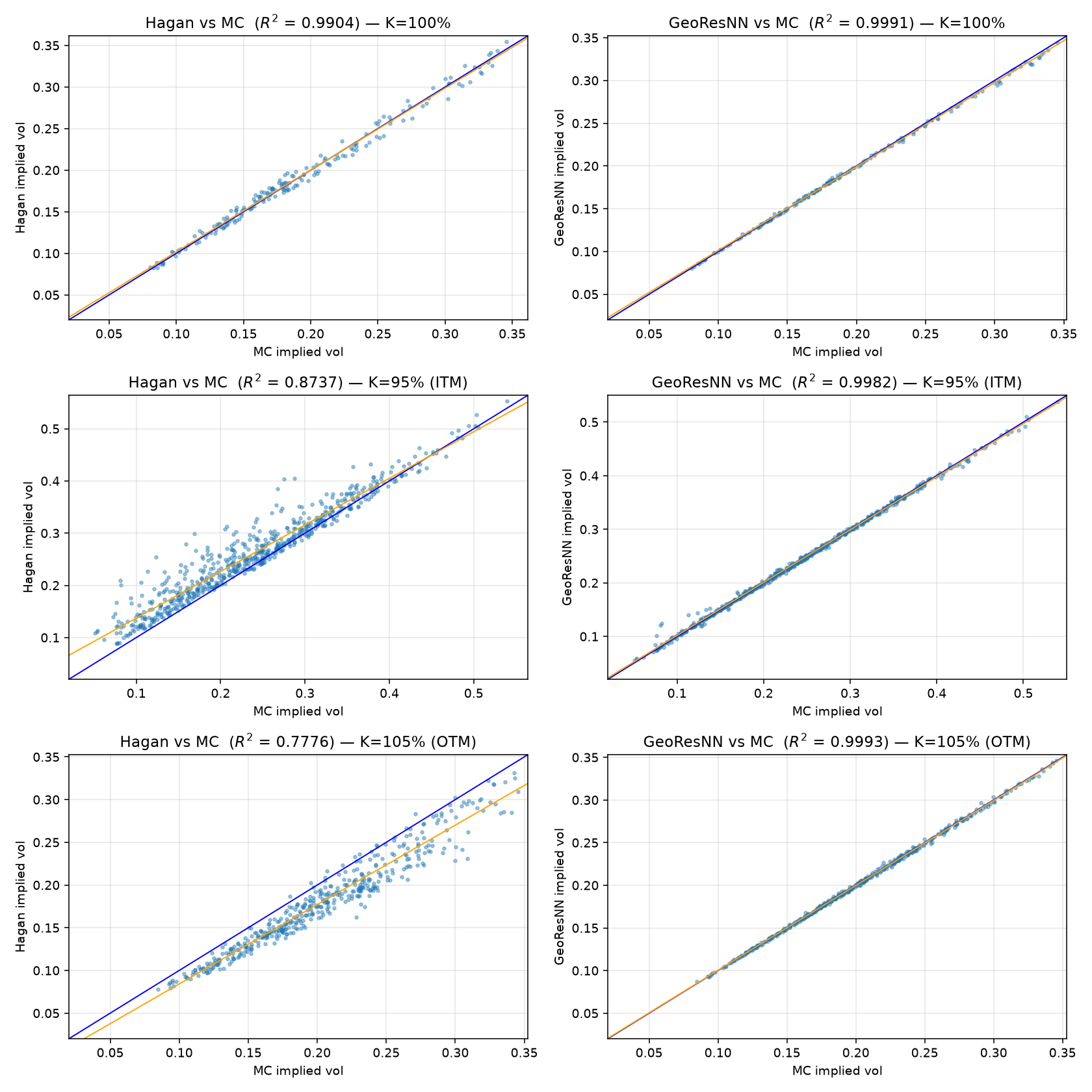}
  \caption{\textbf{GeoResNN correlation diagnostics.}}
  \label{fig:geores2}
\end{figure}
\begin{figure}[!h]
  \centering
  \includegraphics[width=0.9\linewidth]{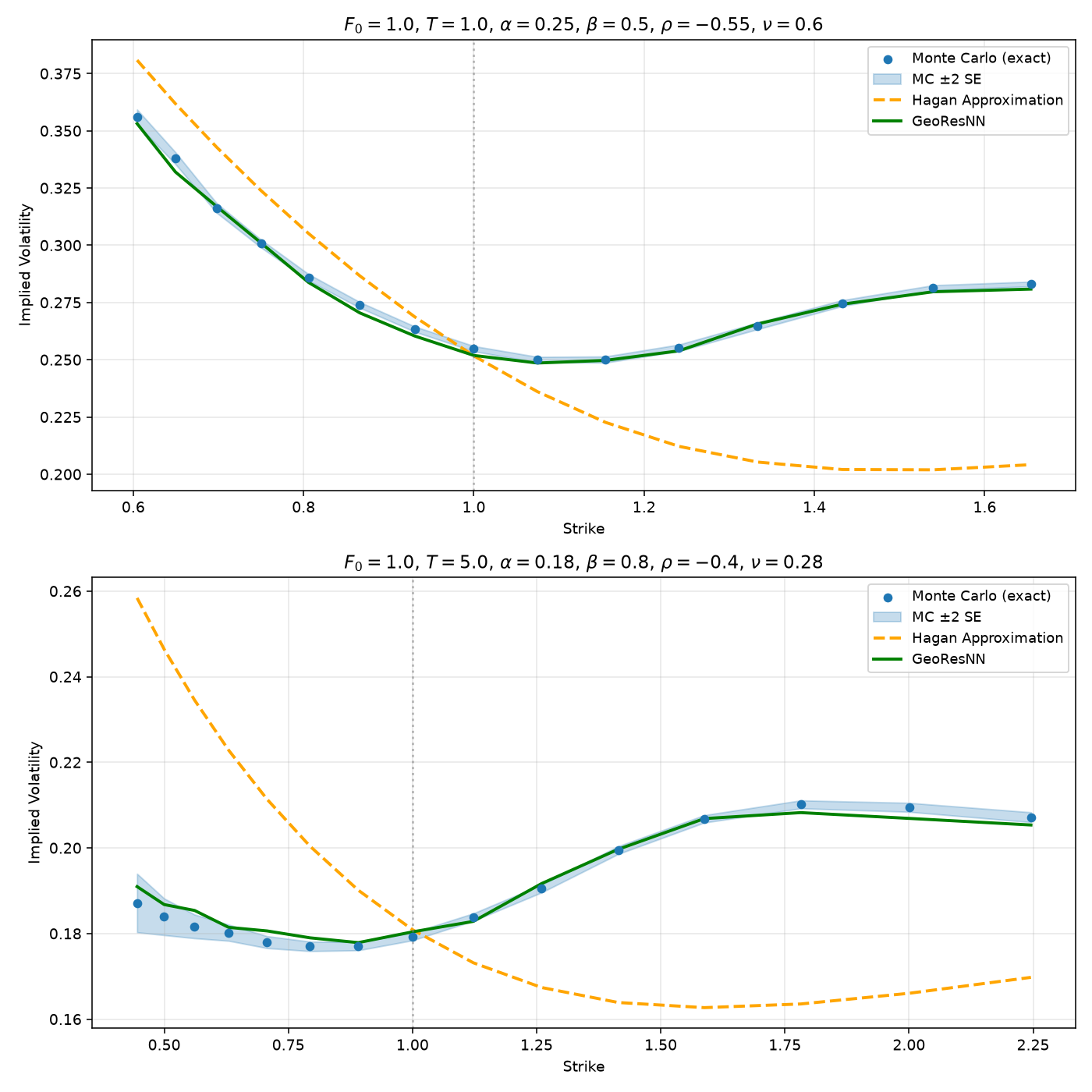}
  \caption{\textbf{GeoResNN under stress.}}
  \label{fig:geores3}
\end{figure}
\begin{figure}[!h]
  \centering
  \includegraphics[width=0.9\linewidth]{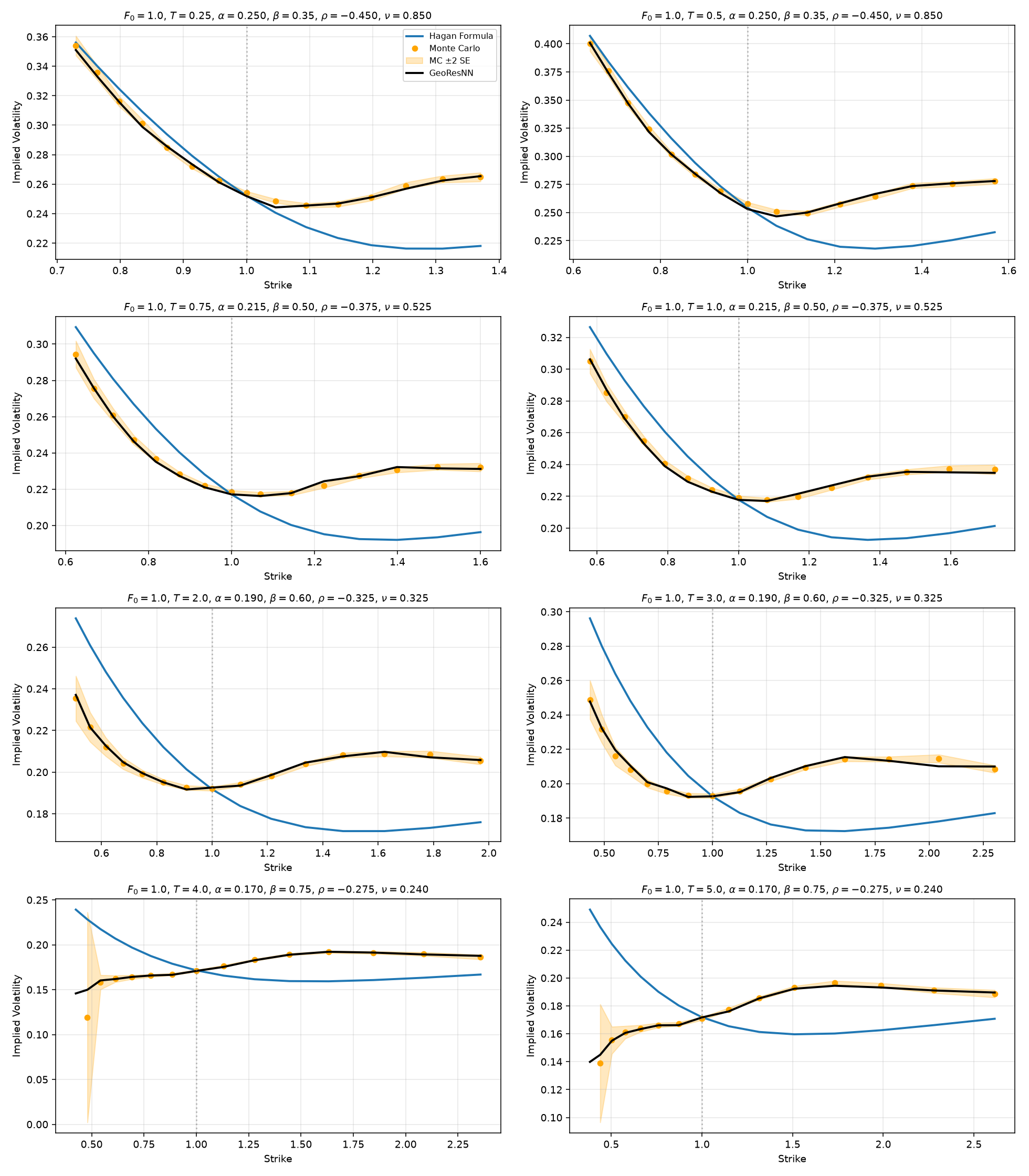}
  \caption{\textbf{GeoResNN smile slices across maturities.}}
  \label{fig:geores4}
\end{figure}

\clearpage
\FloatBarrier

\section*{Funding}
Adil Reghai acknowledges financial support from the Abu Dhabi Investment Authority (ADIA)
through a PhD sponsorship. G\'erard Biau is a member of the Institut Universitaire de France.
No additional external funding was received for this research.

\section*{Author contributions}
Conceptualization: Adil Reghai, G\'erard Biau and Alex Lipton. Methodology: Adil Reghai and
Lama Tarsissi. Formal analysis: Adil Reghai. Software and numerical experiments: Adil Reghai.
Writing -- original draft: Adil Reghai. Writing -- review and editing: Lama Tarsissi,
G\'erard Biau and Alex Lipton. All authors reviewed and approved the final manuscript.

\section*{Conflict of interest}
The authors declare that they have no conflict of interest.

\section*{Data availability}
The synthetic data used in this study were generated through Monte Carlo simulations of the
SABR model. The code used to generate the data and reproduce the numerical experiments is
available from the authors upon reasonable request.

\section*{Acknowledgments}
\added{The authors thank the two anonymous referees for their careful and constructive
reports, which substantially improved the paper. We also thank colleagues and
collaborators for helpful discussions on volatility modeling and machine learning in
quantitative finance.}

\bibliographystyle{apalike}
\bibliography{reference_sabr}

\end{document}